\documentclass[pra,aps,onecolumn,floatfix,superscriptaddress]{revtex4-2}
\usepackage{amsmath}
\usepackage{amssymb}
\usepackage{graphicx}
\usepackage{color}
\usepackage{amssymb}
\usepackage{physics}
\usepackage{soul}
\usepackage[utf8]{inputenc}
\usepackage[T1]{fontenc}
\usepackage{array}
\usepackage{ulem}
\begin{document}

\title{Hybrid Qubit–Qutrit Quantum Battery: Nonclassicality and Energy Performance}
\author{G. Sharvan Prakash}
\affiliation{Department of Physics and Nanotechnology, SRM Institute of Science and Technology, Kattankulathur -- 603203, Tamil Nadu, India}

\author{R. Muthuganesan}
\email{Corresponding author: rajendramuthu@gmail.com}
\affiliation{Department of Physics and Nanotechnology, SRM Institute of Science and Technology, Kattankulathur -- 603203, Tamil Nadu, India}


\begin{abstract}
We propose and analyze a hybrid qubit–qutrit quantum battery (QB) based on a mixed spin-1/2 - spin-1 system interacting via an anisotropic Heisenberg exchange coupling in the presence of a homogeneous magnetic field. The nonclassical properties of the system are characterized using the $l_1-$norm of coherence and negativity, which quantify quantum coherence and entanglement, respectively. The performance of the quantum battery is evaluated through key indicators such as ergotropy, power, and capacity. Our results reveal that both ergotropy and power exhibit oscillatory dynamics, while the capacity remains constant over time. We further investigate the influence of system parameters and magnetic field strength on both quantum correlations and battery performance, demonstrating a strong interplay between nonclassicality and energy-storage performance, as characterized by ergotropy, charging power, and capacity. Importantly, we establish a connection between the theoretical model and an experimentally realizable nickel–radical molecular complex, showing that quantum coherence, entanglement, and efficient energy
storage persist even at room temperature. These findings provide a realistic pathway toward the implementation of hybrid qubit–qutrit quantum batteries in solid-state molecular platforms.
\end{abstract}

\maketitle

\section{Introduction}

The rapid development of quantum technologies has opened new avenues for energy storage in physical systems beyond the limits of classical thermodynamic devices. In recent times, quantum batteries have attracted significant attention as finite-dimensional quantum systems capable of storing and delivering energy by exploiting uniquely quantum mechanical resources such as coherence, entanglement, and collective many-body effects \cite{Oqs,QTD,QTD2,QTD3}. Unlike classical batteries, where energy storage is governed purely by chemical processes, quantum batteries operate within the framework of quantum thermodynamics and quantum resource theory, enabling enhanced charging speed, improved work extraction, and potentially higher power density. These advantages position quantum batteries as promising candidates for next-generation energy-storage architectures.

Over the past decade, quantum batteries have revolutionized the field of Quantum information theory as well as quantum thermodynamics \cite{QTD,QTD2,QTD3,Nielsen2010,QRT1}. Quantum batteries provide a promising framework for quantum energy
storage, with potential advantages in charging power and extractable work. The QBs have been studied experimentally in several studies to determine the quantum advantages during practical implementation. The following are some of the important results that explore the physical realization of a QB. For instance, Florian Metzler et al. (2023) demonstrated a QB based on an organic semiconductor ensemble of two-level systems coupled to a microcavity, achieving a power density up to 672 kW/kg, exceeding that of classical batteries  \cite{DickeQB1,DickeQB2}. Other implementations include nuclear spin systems \cite{NSS}, superconducting circuits \cite{SQB,SQB2}, photonic systems utilizing single-photon polarization \cite{Photon2,Photon3}, and Sachdev–Ye–Kitaev (SYK) battery models describing strongly interacting systems with nonlocal correlations and fast thermalization properties \cite{SYK}. Traditionally, most QB models are constructed using qubits arranged in bipartite or tripartite configurations \cite{Tripartite,Tavis,DipB}. However, recent developments have extended this framework to higher-dimensional systems, including multimode resonators coupled to superconducting transmon qutrits, where advanced charging protocols are actively investigated \cite{Qutrit1,Qutrit2,TQB}. Additionally hybrid cavity magnochemical systems provides an efficient framework for quantum batteries in which the system is composed of  microwave cavity mode, a magnon mode hosted within a YIG-sphere, and a mechanical phonon mode of the sphere’s vibration \cite{MCQB}. Also recent studies on the multi level atomic systems contributes to a better construction of quantum battery because of its large number of excitation pathways supported by their expanded Hilbert spaces \cite{HDQB}. However, classical battery implementations also encompass devices based on a variety of materials \cite{Hu2025,Yang2023,Ye2025,Zhang2025,Gan2022,Wang2025a}.

Among various QB architectures, spin-chain quantum batteries have attracted considerable attention due to their deep connection with condensed matter physics and quantum magnetism. These models, widely studied in quantum many-body systems \cite{SpinModels,1/2spin}, provide a natural platform for energy storage through spin–spin interactions. Unlike cavity-based or photonic batteries, where energy is stored primarily in field modes, spin-chain batteries allow energy to be stored directly in the exchange interactions between spins. In this work, we propose a hybrid spin-chain quantum battery composed of a qubit (spin-1/2) and a qutrit (spin-1) coupled via Heisenberg exchange interaction. This mixed-spin configuration provides a richer Hilbert-space structure and enables the investigation of both nonclassical correlations and performance enhancement in a compact solid-state framework. Notable solid state models considered for the realization of QBs are spin systems where qubits, described as quantum spins, interact under different coupling mechanisms. The spin chains have remarkable contributions in both theoretical and experimental computation because of their integrability and scalability. There are various types of spin chain models including Ising, Heisenberg, Kitaev, etc. Among these Heisenberg spin chains are the most significant and widely studied model \cite{Quspin,HSC}.  The interaction between particles is then handled via perturbation theory or using a mean-field approach where a quadratic Hamiltonian replaces the true Hamiltonian of the system with renormalized temperature-dependent parameters \cite{Spina,SpinModels,1/2spin}.  The charging protocols of the quantum battery determine how energy is injected into the system through controlled unitary or measurement-based operations \cite{chrg1,chrg2}. There are also advanced charging protocols, including measurement-based charging, cavity-assisted charging, and coherently driven charging \cite{chrg3,chrg4,chrg5,Coh}.

Hybrid quantum systems combining subsystems of different spin dimensions offer a particularly attractive platform for quantum energy storage. Such systems naturally possess richer energy spectra, enhanced controllability, and a broader resource structure compared to purely qubit batteries. In recent years, the behavior of thermal entanglement and quantum correlations has been extensively studied \cite{JMM,Venkat}. However, a systematic investigation linking nonclassical correlations—specifically quantum coherence and entanglement—with performance indicators such as ergotropy, charging power, and capacity in hybrid architectures is still lacking. Furthermore, many theoretical quantum battery proposals remain disconnected from experimentally realizable platforms, limiting their relevance for practical implementation.


In this work, we address these gaps by proposing and analyzing a qubit–qutrit quantum battery modeled as a mixed spin-(1/2, 1) Heisenberg dimer subjected to an external magnetic field. The hybrid spin architecture enables simultaneous exploration of quantum resource generation and energy-storage performance within a unified theoretical framework \cite{JMM,QB,DMI,QBDM}. We quantify nonclassicality using the $l_1$-norm of coherence and negativity, while battery performance is characterized through ergotropy, instantaneous charging power, and energy capacity. The interplay between these quantities reveals how quantum
correlations are associated with enhanced ergotropy and charging power during the cyclic unitary evolution.  Importantly, our model is not merely abstract but admits a realistic physical realization in nickel–radical molecular magnetic complexes, which naturally implement a mixed spin-(1/2, 1) system with strong exchange coupling and experimentally controllable magnetic driving. By mapping our theoretical parameters onto experimentally reported values, we demonstrate that the predicted oscillatory charging dynamics, coherence generation, and entanglement persist under accessible laboratory conditions, including near room temperature \cite{Nickel2, Nickel3,Nickel}. This establishes a concrete pathway toward experimental realization of hybrid quantum batteries in molecular spin platforms.

The remainder of the paper is organized as follows. In Sec. II, we introduce the Hamiltonian of the qubit–qutrit quantum battery and describe the charging protocol. Section III defines the quantum correlation measures and performance indicators. In Sec. IV, we present and analyze the dynamical results for coherence, entanglement, ergotropy, and power under varying system parameters. Section V discusses the experimental relevance and feasibility in molecular magnetic systems. Finally, Sec. VI summarizes our conclusions and outlines future directions for hybrid quantum energy-storage devices.
\section{Quantum Battery}

\begin{figure}[!h]
    \begin{center}
    \includegraphics[width=0.5\linewidth]{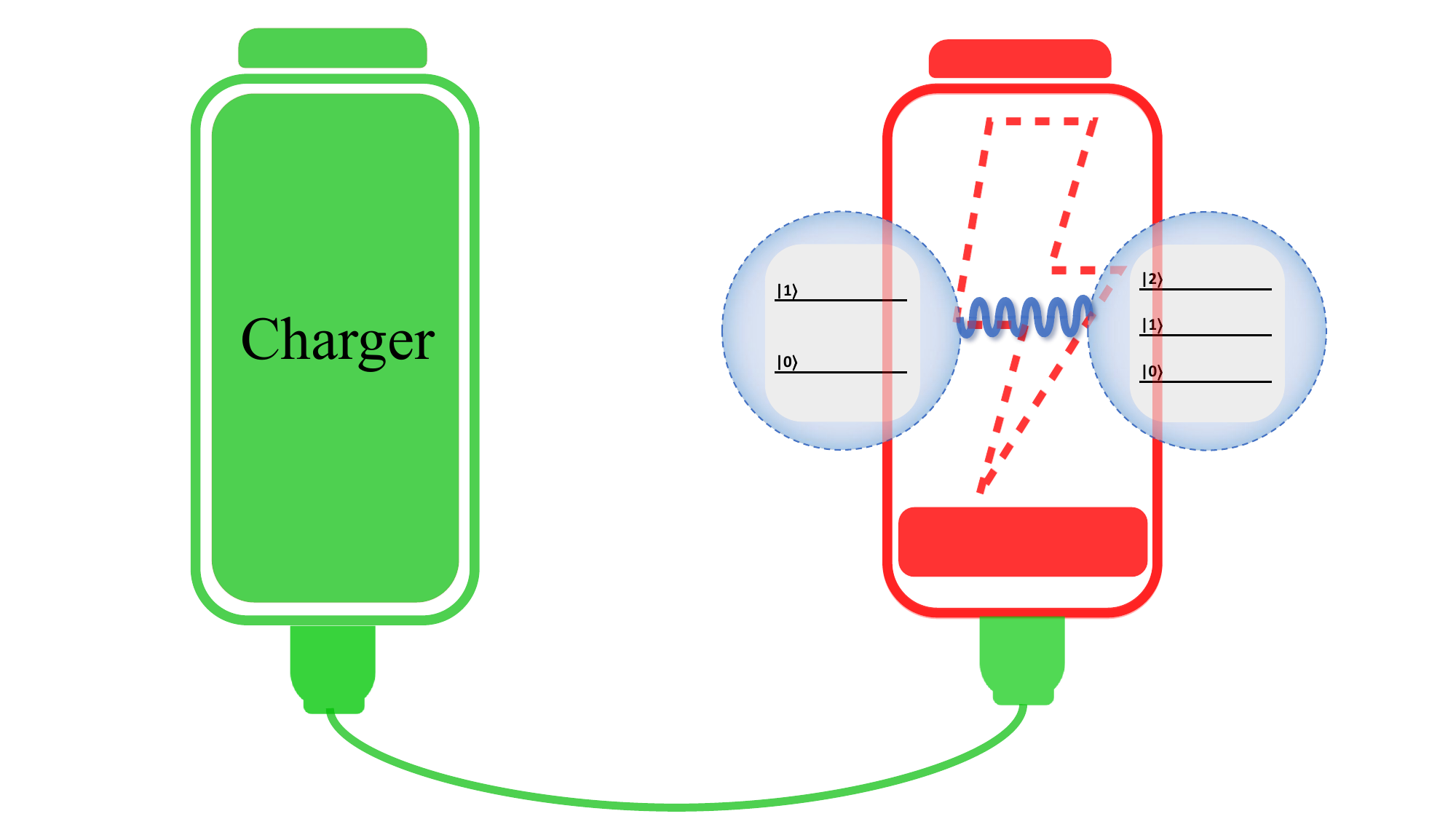}
    \caption{Schematic representation of Quantum battery in Qubit-Qutrit system}
    \label{model}
    \end{center}
\end{figure}

In this section, we introduce the theoretical Hamiltonian of the hybrid quantum battery composed of a qubit and a qutrit which is represented in Fig.({\ref{model}}). The spin-1/2 and spin-1 systems are coupled via an anisotropic exchange interaction in the presence of a homogeneous magnetic field.
\subsection{ Hamiltonian of the battery}
To analyze the nonclassicality and performance of a qubit–qutrit quantum battery, we consider a hybrid spin system consisting of a spin-1/2 (qubit) and a spin-1 (qutrit) interacting via an anisotropic Heisenberg exchange interaction in the presence of uniaxial single-ion anisotropy and a homogeneous magnetic field. The Hamiltonian of the quantum battery is given by \cite{JMM, Dimer}
\begin{align}
\mathcal{{H}_{B}}=J\left[\Delta\left(\hat{s}^{x}\hat{S}^{x}+\hat{s}^{y}\hat{S}^{y}\right)+\hat{s}^{z}\hat{S}^{z}\right]
+D(\hat{S}^{z})^{2}
-g_{1}\mu_{B}B\hat{s}^{z}
-g_{2}\mu_{B}B\hat{S}^{z},
\label{Battery}
\end{align}
where $\hat{s}^{\alpha}$ ($\hat{S}^{\alpha}$) denotes the spatial components of the spin-$1/2$ (spin-$1$) operators with $\alpha=x,y,z$. Here, $J$ denotes the exchange coupling constant between the spin-$1/2$ and spin-$1$ particles, $\Delta$ is the exchange anisotropy parameter, and $D$ represents the uniaxial single-ion anisotropy of the spin-$1$ particle \cite{JMM, Dimer}. The parameters $g_{1}$ and $g_{2}$ are the Land\a 'e $g$-factors corresponding to the spin-$1/2$ and spin-$1$ magnetic ions, respectively, while $B$ denotes the strength of the homogeneous external magnetic field.


In the standard computational basis $\{ |00\rangle, |01\rangle, |02\rangle, |10\rangle, |11\rangle,|12\rangle \}$ of the qubit–qutrit system, the eigenvalues and the corresponding eigenvectors of the Hamiltonian in Eq.~(\ref{Battery}) are computed as follows:

\begin{align}
    \lambda_{1,2}&=\frac{1}{2} \left(J+2D\mp(h_1+2h_2)\right), ~~~~~~~ |\varphi_1\rangle=|\frac{1}{2},1\rangle, ~~|\varphi_2\rangle=|-\frac{1}{2},1\rangle, \nonumber \\
    \lambda_{3,4}&=\frac{1}{4}\Lambda_+ \mp \eta_{-}, ~~~~~~~~~~~~~~~~~~~~~~~~~~ |\varphi_{3,4}\rangle= \delta_{\mp}|\frac{1}{2},0\rangle\mp \delta_{\pm} |-\frac{1}{2},1\rangle, \nonumber \\
     \lambda_{5,6}&=\frac{1}{4}\Lambda_-\mp \eta_{+}, ~~~~~~~~~~~~~~~~~~~~~~~~~~ |\varphi_{5,6}\rangle=\chi_{\pm} |\frac{1}{2},-1\rangle \mp \chi_{\mp}|-\frac{1}{2},0\rangle, \nonumber
\end{align}
where $h_{j}=g_j\mu_B B$ , $\eta_{\pm}=\sqrt{[J-2D \pm 2(h_1 - h_2)}]^2 +8(J \Delta)^2$ and the normalization constants are given by$\\
    \mathcal{\delta}_{\pm}= \frac{1}{\sqrt{2}}\sqrt{1\pm \frac{\eta_-}{\eta^2_-+8J^2}}, ~~\chi_{\pm}=\frac{1}{\sqrt{2}}\sqrt{1\pm \frac{\eta_+}{\eta^2_++8J^2}}$ and $\Lambda_{\pm}=-2D\pm 2h_2+J$.\\
    
For clarity, we introduce a common reference energy scale $E_0$ and express all energetic parameters in dimensionless form. We define
\begin{equation}
\tilde{J}=\frac{J}{E_0},\qquad
\tilde{D}=\frac{D}{E_0},\qquad
\tilde{\Omega}=\frac{\hbar\Omega}{E_0},\qquad
\tilde{h}_i=\frac{g_i\mu_B B}{E_0},
\end{equation}
with the dimensionless temperature and time
\begin{equation}
\tilde{T}=\frac{k_B T_{\rm phys}}{E_0},
\qquad
\tau=\frac{E_0t_{\rm phys}}{\hbar}.
\end{equation}
Thus, all energies appearing in the dimensionless Hamiltonian are measured in units of $E_0$, while time is measured in units of $\hbar/E_0$. Accordingly, the ergotropy $W$ and capacity $K$ are reported in units of $E_0$, whereas the charging power $P=dW/dt$ is reported in units of $E_0^2/\hbar$. The quantities $\Delta$, $g_1$, and $g_2$ are dimensionless. Natural units $\hbar=k_B=\mu_B=1$ are used only in the dimensionless calculations. Whenever the magnetic field is quoted in tesla, the physical Bohr magneton is restored explicitly and the Zeeman coupling entering the dimensionless Hamiltonian is $\tilde{h}_i=g_i\mu_B B/J$.

The thermal density matrix of the system $\mathcal{H}_B$ is defined as,
\begin{align}
\varrho(0,\tilde{T})=\frac{1}{\mathcal{Z}}\text{exp}(-\beta \mathcal{H}_{B})=\frac{1} {\mathcal{Z}}\sum\limits_{i=1}^{6} p_{i}|\varphi_{i}\rangle \langle \varphi_{i}|, 
\end{align}
\label{Thermaldef}
where $\beta=1/k_{B}\tilde{T}$ and $p_i$ are the eigenvalues of the state $\varrho(0,\tilde{T})$. For the sake of simplicity $k_B$ set to unity. The explicit matrix elements of the thermal state $\varrho(0,\tilde{T})$ are presented in Appendix~\ref{A}.

\subsection{Charging Protocol}
The charging process of the qubit–qutrit quantum battery is modeled by an external driving Hamiltonian of the form \cite{SQB, OQB, Erg, Erg1,Erg2,chrg1,chrg2,Bloch}
\begin{align}
\mathcal{H}_{c}
=\Omega\left[\cos{\theta}(\hat s^{z}\otimes\mathbb{I}_{3})
+\sin{\theta}(\mathbb{I}_{2}\otimes \hat S^{z})\right],
\label{charger}
\end{align}
where $\Omega$ denotes the overall charging strength that sets the energy scale and rate of the charging protocol. The parameter $\theta$ controls the relative contribution of the local driving fields acting on the qubit and qutrit subsystems, respectively. 
Here, $\hat s^{z}$ and $\hat S^{z}$ are the $z$-components of the spin operators for the spin-$1/2$ (qubit) and spin-$1$ (qutrit), while $\mathbb{I}_{2}$ and $\mathbb{I}_{3}$ represent the identity operators in the qubit and qutrit Hilbert spaces. The first term in Eq.~(\ref{charger}) corresponds to a local longitudinal driving field applied to the qubit, whereas the second term represents an analogous driving field acting on the qutrit.

By tuning the angle $\theta$, one can continuously interpolate between purely qubit charging ($\theta=0$), purely qutrit charging ($\theta=\pi/2$), and hybrid charging regimes where both subsystems are driven simultaneously. This flexibility enables a systematic investigation of how local versus distributed charging strategies influence the energy extraction, storage capacity, and nonclassical correlations of the hybrid quantum battery. The time evolution of the density matrix is given by
\begin{align}
    \varrho(\tau,\tilde{T})=\mathcal{U}(\tau)\varrho(0,\tilde{T})\mathcal{U}(\tau)^{\dagger}
\end{align}
where $\mathcal{U}(\tau)=\exp^{-i\mathcal{H}_{c}\tau}$ is the unitary operator. The matrix elements of the time-evolved quantum battery are given in the Appendix. \ref{evoleved}. In realistic molecular systems, environmental interactions such as spin–phonon coupling and thermal noise lead to decoherence and relaxation effects. These processes can be modeled within the Lindblad master equation framework. In the presence of dephasing and dissipative channels, quantum coherence and entanglement are expected to decay exponentially with time, leading to a suppression of ergotropy and charging power.

Nevertheless, the qualitative behavior observed in the present study remains robust, as it is governed by intrinsic system parameters. In particular, molecular magnetic systems are known to exhibit relatively long coherence times, allowing the essential features of quantum-enhanced energy storage to persist within experimentally accessible regimes.
\section{Observables}
In this section, we review the quantum correlation measures, including the $l_1-$norm of coherence and negativity, as well as the performance indicators of the quantum battery, namely ergotropy, power, and capacity. Let $\varrho$ be the bipartite density matrix representing the state of the quantum battery in the separable Hilbert space $H^2\otimes H^3$. 
\subsection{Quantum Correlation Measures}
\textit{\bf Negativity:}
The quantification of entanglement in higher-dimensional systems is a central problem in quantum information theory.  For $2\otimes n$ systems, negativity is particularly valuable because it provides a computable and operationally meaningful quantifier of entanglement, making it suitable for analyzing hybrid systems such as qubit–qutrit quantum batteries \cite{Neg,Neg2,Neg3}. It is based on the positive partial transpose (PPT) criterion, which states that a bipartite quantum state $\varrho$ is separable if and only if its partial transpose is positive semidefinite for $\mathrm{2}\otimes \mathrm{3}$ systems.
 For a general $\mathrm{2}\otimes \mathrm{3}$ bipartite state $\varrho$, the negativity is defined as \cite{Neg4}
\begin{align}
    \mathcal{N}(\varrho)=\frac{||\varrho^{T_A}||_1-1}{2},
\end{align}
where $\varrho^{T_A}$ denotes the partial transpose of $\varrho$ with respect to the marginal system $A$ and $||X||_1=\text{Tr}\sqrt{X^{\dagger}X}$ is the trace norm. Equivalently, negativity can be expressed as
\begin{align}
    \mathcal{N}(\varrho)=\sum_{\lambda_i < 0}|\lambda_i| 
\end{align}
where $\lambda_i$ are the eigenvalues of the partially transposed density matrix $\varrho^{T_A}$. A nonzero value of $\mathcal{N}(\varrho)$ indicates the presence of entanglement between the subsystems, whereas $\mathcal{N}(\varrho)=0$ for seperable state.

\textit{\bf $l_1-$norm of coherence:}
Quantum coherence is a fundamental resource that captures the superposition principle of quantum mechanics and plays a crucial role in quantum technologies, including quantum thermodynamic devices and quantum batteries. Within the framework of resource theory, coherence is defined with respect to a fixed reference basis $\{ |i\rangle \}$, and several quantifiers have been proposed to characterize its magnitude.

Among them, the $l_{1}$-norm of coherence is one of the most widely adopted measures due to its simplicity and clear physical interpretation. For a given density matrix $\varrho$ expressed in the reference computational basis $\{ |i\rangle \}$, the $l_{1}$-norm of coherence is defined as \cite{Baumgratz2014,Streltsov2017}
\begin{align}
    C_{l_{1}}(\varrho) = \sum_{i \neq j} |\varrho_{ij}^\tau|,
\end{align}
where $\varrho_{ij}^{\tau} = \langle i | \varrho | j \rangle$ are the off-diagonal elements of the density matrix.

This measure directly quantifies the total amount of quantum superposition present in the system, since only off-diagonal terms contribute to coherence. A vanishing value $C_{l_{1}}(\varrho)=0$ indicates an incoherent (classical) state that is diagonal in the chosen basis, while a nonzero value signals the presence of genuine quantum coherence. Owing to its operational significance and computational simplicity, the $l_1$-norm of coherence provides a useful tool for analyzing the relationship between quantum coherence and battery-performance indicators such as extractable work and charging dynamics.

\subsection{Ergotropy and Power}
\textit{\bf Ergotropy:}  One of the important quantity which governs the performance of a quantum battery is called the Ergotropy $\mathcal{W}(\tau)$. This quantifies the amount of extractable work from the QB through a cyclic unitary operations after the charging protocol \cite{OQB,Erg,Erg1,Erg2,Erg3}.
The ergotropy is calculated by performing an optimization over all possible unitary operations to achieve a final state that has minimum average energy with respect to $\mathcal{H_B}$. After extracting the maximum amount of work during the unitary process the system reaches its passive state which is diagonal in the Hamiltonian's eigen basis. Also when the battery's final state is passive no more work can be extracted under unitary operations. Ergotropy is generally defined by the formula as follows;
\begin{align}
     \mathcal{W}(\tau)=\text{Tr}\{\mathcal{H}_{B}\varrho(\tau,\tilde{T})\}-\min_{U\epsilon\hat{\mathcal{U}}}\{\text{Tr}\{\mathcal{H}_{B}U\varrho U^\dagger\}\}
\end{align}
where $\mathcal{\hat{U}}$ is the set of all accessible unitary evolutions, and the minimum energy is obtained by the unitary evolutions. 
\begin{align}
    \mathcal{W}(\tau)=\text{Tr}[\mathcal{H}_{B}\varrho(\tau,\tilde{T})]-\text{Tr}[\mathcal{H}_{B}\varrho_{passive}(\tau)]
\end{align}
where $\varrho(\tau,\tilde{T})$ is the time-evolved thermal density matrix, $\varrho_{passive}(\tau)$ is the passive counterpart of $\varrho(\tau,\tilde{T})$ with respect to Hamiltonian of the QB $\mathcal{H}_{B}$. 
The passive counterpart is defined by\\
\begin{align}
   \varrho_{passive}(\tau)=\sum_i {p_i}^\downarrow |{{E_i}^\uparrow}\rangle \langle{{E_i}^\uparrow}|,
\end{align}
where ${p_i}^\downarrow$ is the eigenvalues of $\varrho(\tau,\tilde{T})$ sorted in descending order, and $|{{E_i}^\uparrow}\rangle$ are the eigenstates of $\mathcal{H}_{B}$ arranged in ascending order of energy.\\
\textit{\bf Power:} Power $\mathcal{P}(\tau)$ is also an another quantity that evaluates the performance of the QB. The instantaneous power of the QB is the time derivative of the ergotropy $\mathcal{W}(\tau)$ i.e., rate at which extractable work is stored in the battery during the charging protocol. It is defined by;
\begin{align}
    \mathcal{P}(\tau)=\frac{d\mathcal{W}(\tau)}{d\tau}.
\end{align}
\textit{\bf Capacity:} The capacity $\mathcal{K}$ of the quantum battery is given by;
\begin{align}
    \mathcal{K}=Tr[\mathcal{H}_{B} \hat{\varrho}_{\uparrow}]-Tr[ \mathcal{H}_B\hat{\varrho}_{\downarrow}]
\end{align}
where $\hat{\varrho}_{\uparrow}=|1^{\otimes N}\rangle\langle1^{\otimes N}|$ and $\hat{\varrho}_{\downarrow}=|0^{\otimes N}\rangle\langle0^{\otimes N}|$ represents the excited and ground states of the N-partite quantum battery respectively. This measure does not necessarily quantifies the performance but the energy gap between the maximum and minimum energy states of the QB. 
\section{Results and Discussion}
\label{sec3}
\begin{figure}[!h]
	\begin{center}
		\includegraphics[width=0.45\textwidth, height=145px]{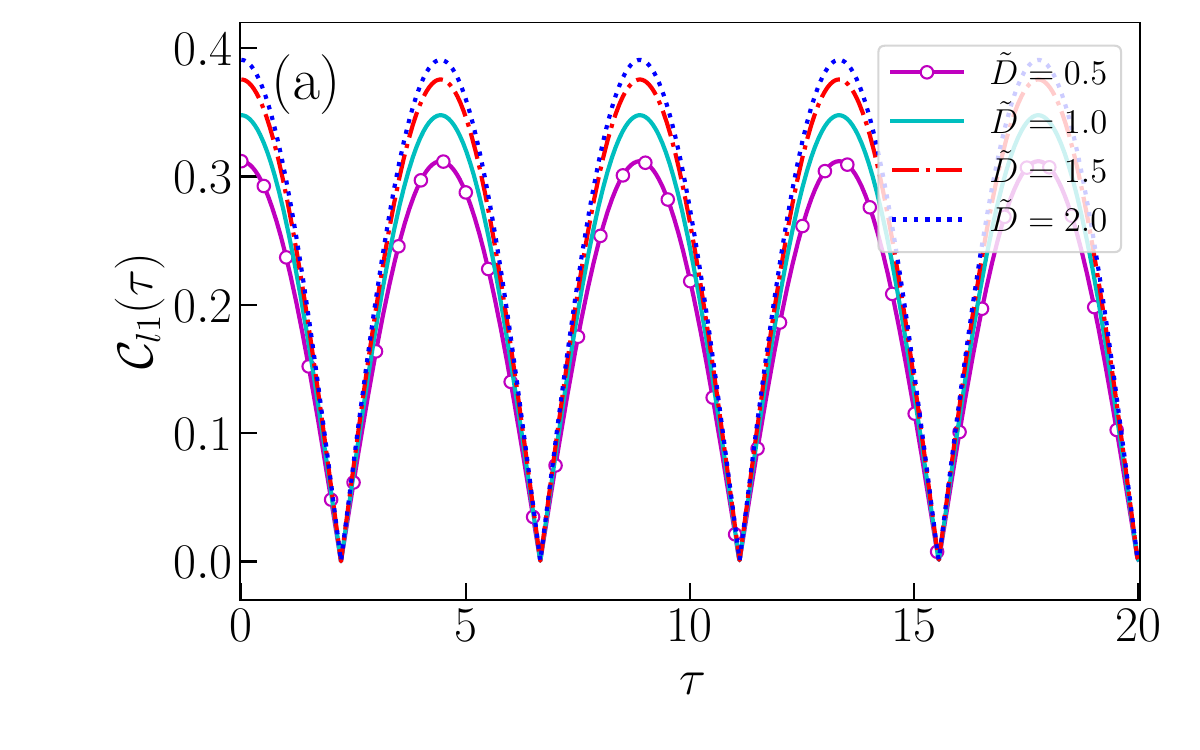}
		\includegraphics[width=0.45\textwidth, height=145px]{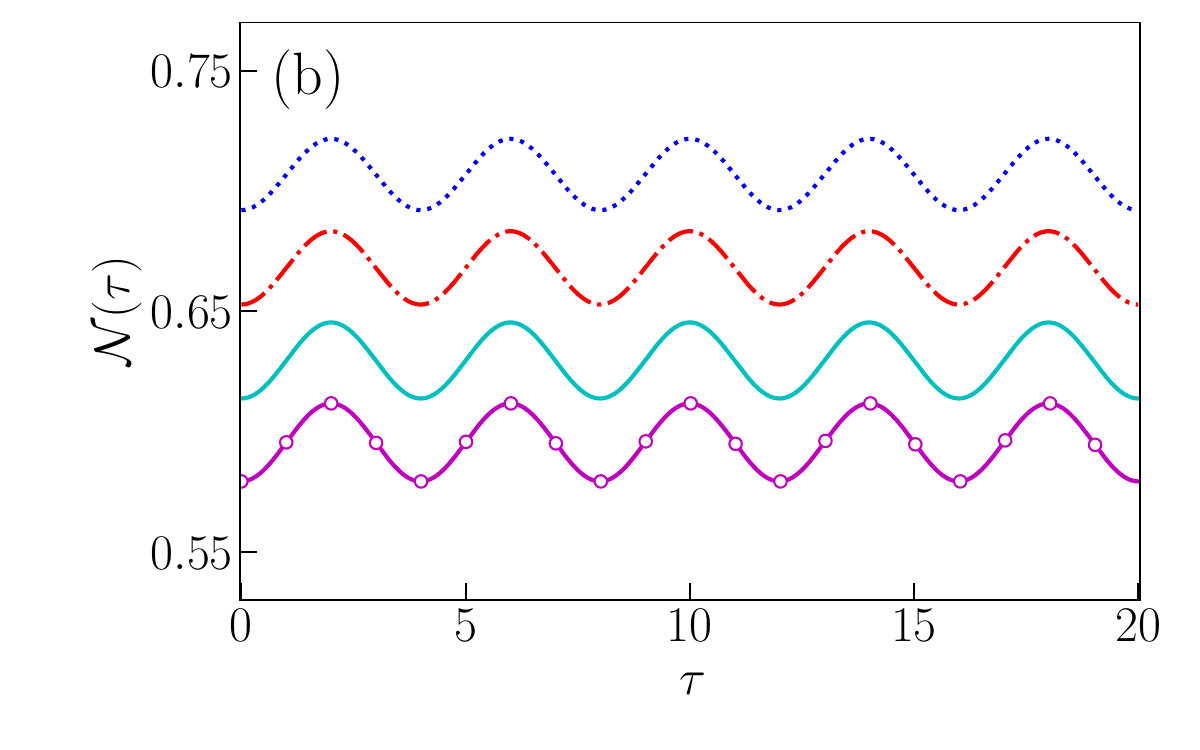}
  		\end{center}
\caption{ Temporal evolution of (a) $l_1-$norm of coherence and (b) negativity of hybrid quantum battery as a function of time $t$ for different values of single ion anisotropy $\tilde{D}$ with the parameters set $\theta=\pi/4$, $g_{1}=g_{2}=2$ and $B=\tilde{J}=\tilde{\Omega}=\tilde{T}=1$.}
\label{fig1}
\end{figure}

In what follows, we investigate the dynamical aspect of quantumness and the performance of qubit-qutrit quantum battery as a function of the dimensionless system's parameters and external magnetic field. The results are obtained under the unitary charging protocol governed by the charging Hamiltonian Eq. (\ref{charger}), with the initial state prepared as a thermal state of the battery Hamiltonian. Fig. (\ref{fig1}) illustrates the temporal evolution of the $l_1$-norm of coherence and negativity of the qubit–qutrit quantum battery as functions of time $t$ for different values of the uniaxial single-ion anisotropy parameter $\tilde{D}$ and other  fixed parameters are $g_1=g_2=2$, $B=1$, and $\tilde{J}=1$, where the time variable is expressed as dimensionless units through the scaling $t \rightarrow \Omega t_{phys}$ with $t_{phys}$ being physical charging time. A dimensionless time $t=1$ corresponds to approximately $0.16ns$ for a typical microwave pulse of driving strength($\Omega/2\pi\approx1$)GHz. The coherence, shown in  Fig. (\ref{fig1})a, exhibits pronounced periodic oscillations arising from the coherent unitary charging protocol. An increase in the anisotropy parameter $\tilde{D}$ leads to a systematic enhancement of the coherence amplitude, indicating that stronger single-ion anisotropy promotes sustained quantum superposition between the qubit and qutrit subsystems. This behavior originates from the anisotropy-induced modification of the energy spectrum, which enhances coherent population mixing during the charging process. The negativity displays a similar oscillatory behavior, reflecting the dynamical generation and modulation of entanglement in the hybrid system. Furthermore, the intervention of $\tilde{D}$ do not alter the frequency of oscillation of quantumness quantifiers.

Notably, higher values of $\tilde{D}$ result in an overall increase in the entanglement magnitude, demonstrating that single-ion anisotropy acts as an effective control knob for strengthening nonclassical correlations in the battery. The concurrent enhancement of coherence and entanglement highlights the constructive role of single-ion anisotropy in engineering quantum resources relevant for energy storage. These results establish a direct link between tunable microscopic interactions and macroscopic battery-relevant properties, underscoring the importance of anisotropy-driven resource control in the design of high-performance hybrid quantum batteries.

\begin{figure}[!h]
	\begin{center}
		\includegraphics[width=0.3\textwidth, height=145px]{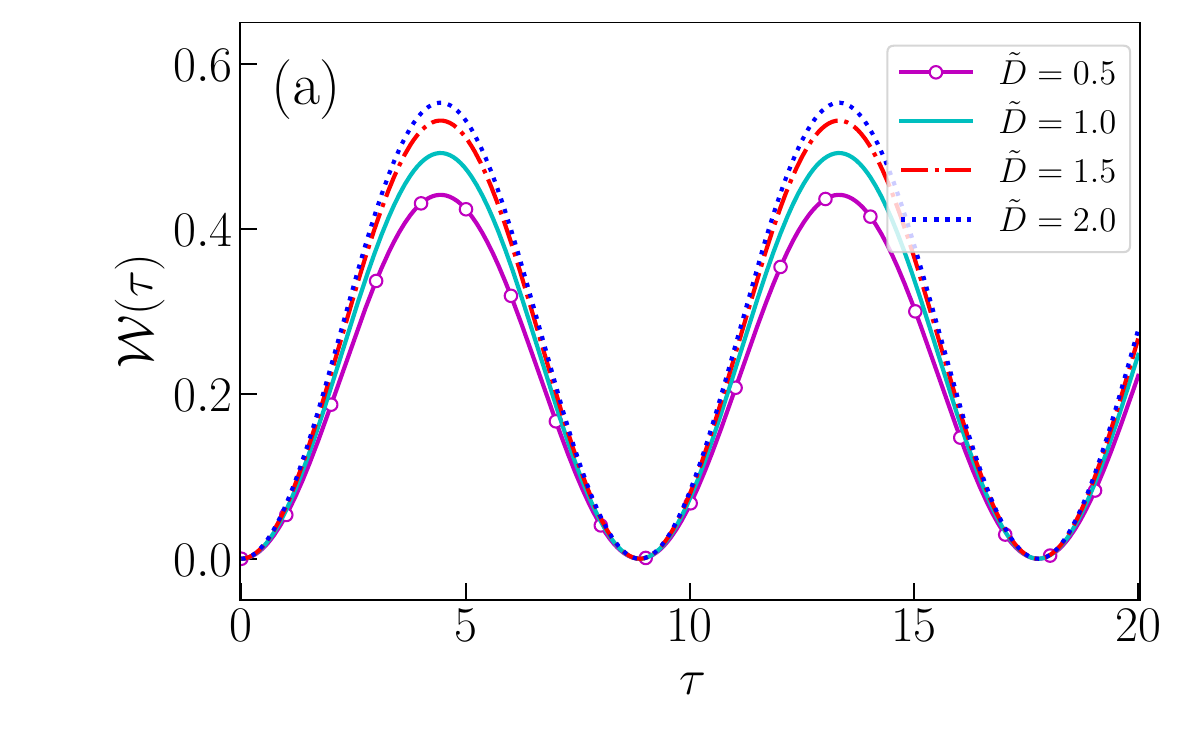}
		\includegraphics[width=0.3\textwidth, height=145px]{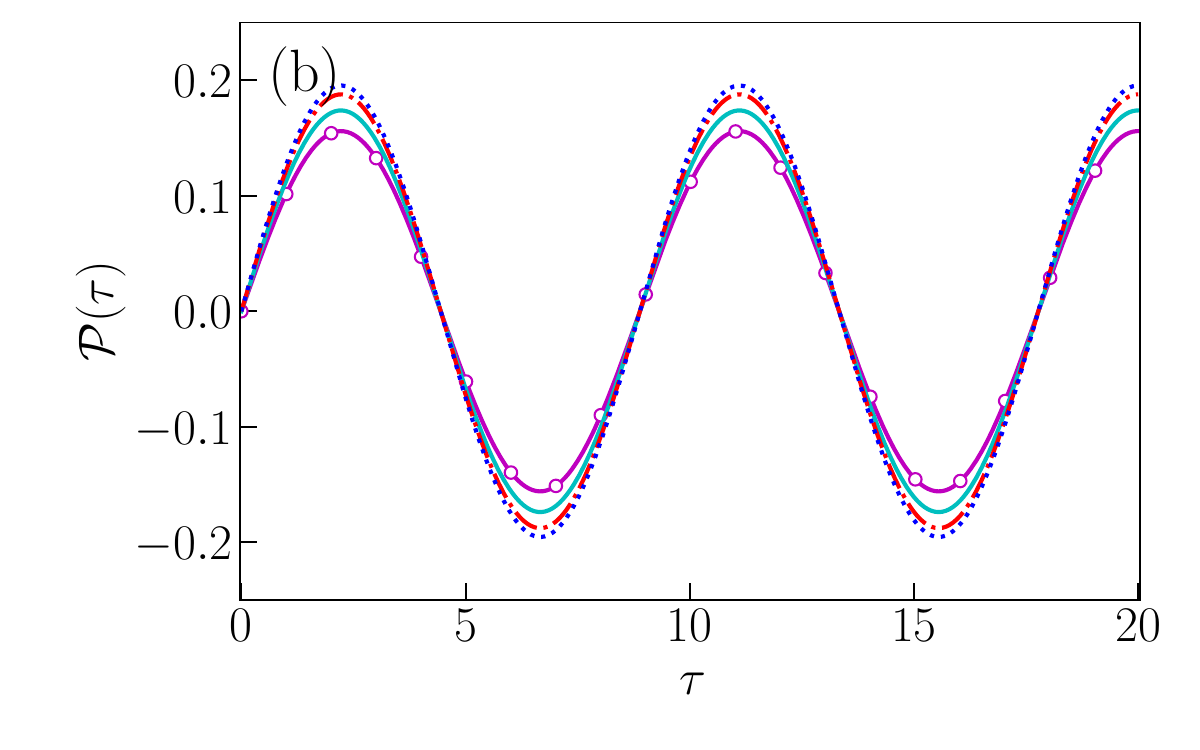}
        \includegraphics[width=0.30\textwidth, height=145px]{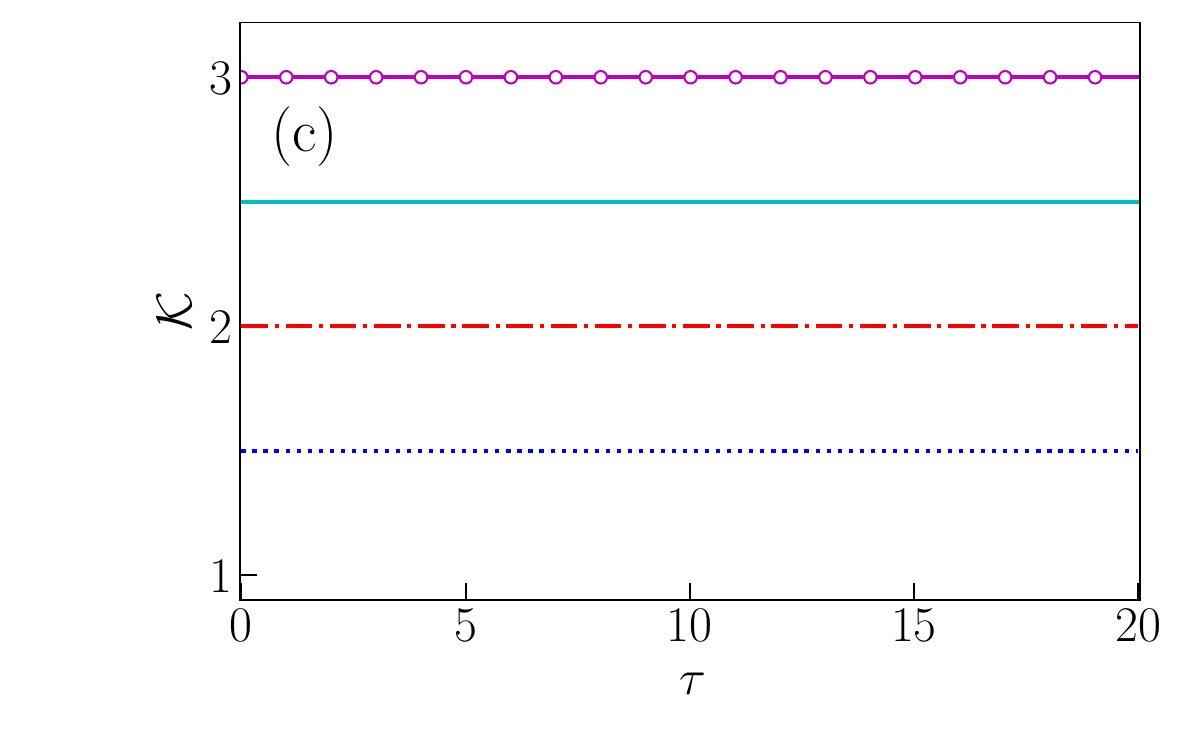}
  		\end{center}
\caption{  Temporal evolution of (a) ergotropy, (b) power and (c) capacity of hybrid quantum battery as a function of time at a few selected values of single ion anisotropy $\tilde{D}$ with the parameters set $\theta=\pi/4$, $g_{1}=g_{2}=2$ and $B=\tilde{J}=\tilde{\Omega}=\tilde{T}=1$.}
\label{fig2}
\end{figure}

Figure \ref{fig2} illustrates the temporal evolution of the key performance indicators of the qubit--qutrit quantum battery, namely ergotropy, charging power, and capacity, for selected values of the single ion anisotropy parameters. The ergotropy quantifies the maximum extractable work from the quantum battery through cyclic unitary operations and serves as a direct measure of the battery usefulness. As shown in Fig. (\ref{fig2}) a, the ergotropy exhibits periodic oscillations in time similar to nonclassical quantifiers, reflecting coherent energy exchange between the charger and the battery. The oscillation amplitude is strongly influenced by the anisotropy parameter, indicating tunable work extraction capability. Also the peak ergotropy exhibits an approximately linear dependence on the anisotropy parameter within the studied parameter range. And the fitting equation is approximately given by $\mathcal{W}(\tilde{D})_{max}$ $\approx 0.075\tilde{D} +0.367 $. In Fig. (\ref{fig2})b, we study the influence of single ion anisotropy parameter on power with respect to time. The charging power, defined as the time derivative of ergotropy, displays alternating positive and negative peaks, corresponding to charging and partial discharging intervals during the cyclic evolution. The maximum power increases in regimes where coherent dynamics dominate, highlighting the role of quantum control in accelerating energy storage. The capacity, which represents the intrinsic energy gap between the maximum and minimum energy eigenstates of the battery Hamiltonian, remains time-independent, as expected. The capacity $\mathcal{K}$ of qubit-qutrit QB is computed as
\begin{align}
    \mathcal{K}=\tilde{h}_1+\tilde{h}_2-\tilde{D} - \frac{\tilde{J}}{2}.
\end{align}
Thus, the capacity $\mathcal{K}$ is a time independent quantity and depends solely on the exchange coupling constant $\tilde{J}$ and the uniaxial single ion anisotropy $\tilde{D}$. In other words, its magnitude depends solely on the static parameters of the Hamiltonian and sets an upper bound on the achievable ergotropy.

A direct comparison between Fig. (\ref{fig1}) and Fig. (\ref{fig2}) reveals a strong correlation between battery performance and nonclassicality. The time intervals in which ergotropy and charging power attain large values are accompanied by enhanced values of the $l_1$-norm of coherence and negativity. This correspondence indicates a close relationship between the evolution of quantum coherence and entanglement and the work-extraction dynamics of the hybrid quantum battery. However, the simultaneous variation of these quantities should not be interpreted as establishing a direct causal relationship between nonclassicality and battery performance. Figure (\ref{fig2}) establishes that the performance of a qubit–qutrit quantum battery is intrinsically linked to its underlying quantum correlations. Enhanced coherence and entanglement are observed to be accompanied by
larger values of ergotropy and charging power under the considered parameter regime,  highlighting the importance of quantum-resource engineering for the development of high-performance quantum energy storage devices.


\begin{figure}[!h]
\begin{center}
        \includegraphics[width=0.45\textwidth, height=145px]{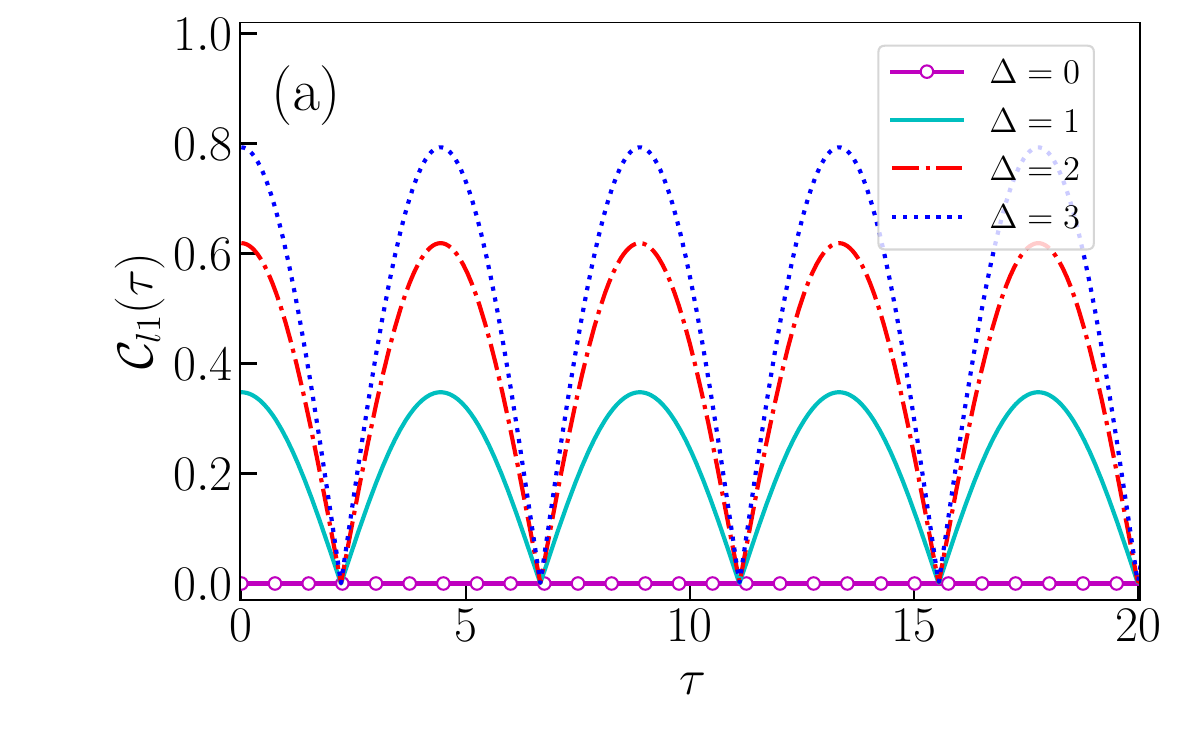}\label{}
        \includegraphics[width=0.45\textwidth, height=145px]{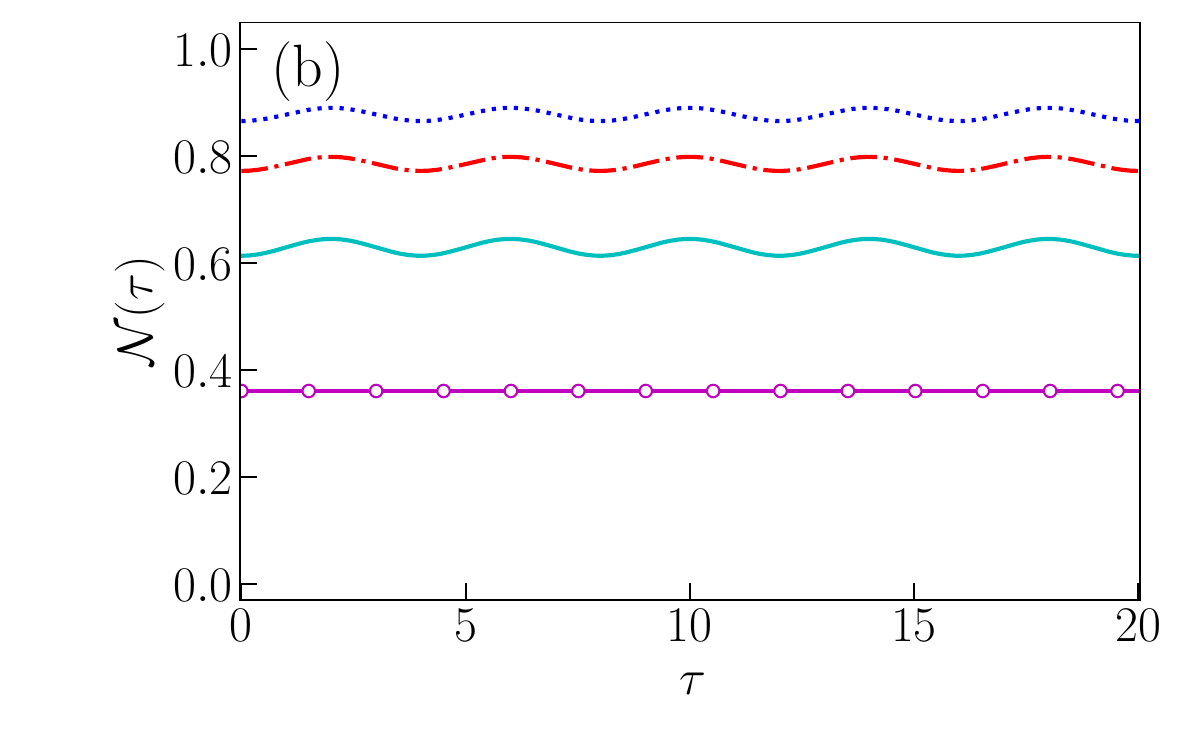}\label{}
        \end{center}
\caption{ Temporal evolution of (a) $l_1-$norm of coherence and (b) negativity of the hybrid quantum battery as a function of time $t$  for different values of the exchange anisotropy parameter $\Delta$, which controls the relative strength of transverse and longitudinal spin interactions. The parameters fixed $\theta=\pi/4$, $g_1=g_2=2$ and $\tilde{J}=\tilde{D}=\tilde{\Omega}=\tilde{T}=1$.} 
\label{del1}
\end{figure}

\begin{figure}[!h]
\begin{center}
        \includegraphics[width=0.3\textwidth, height=120px]{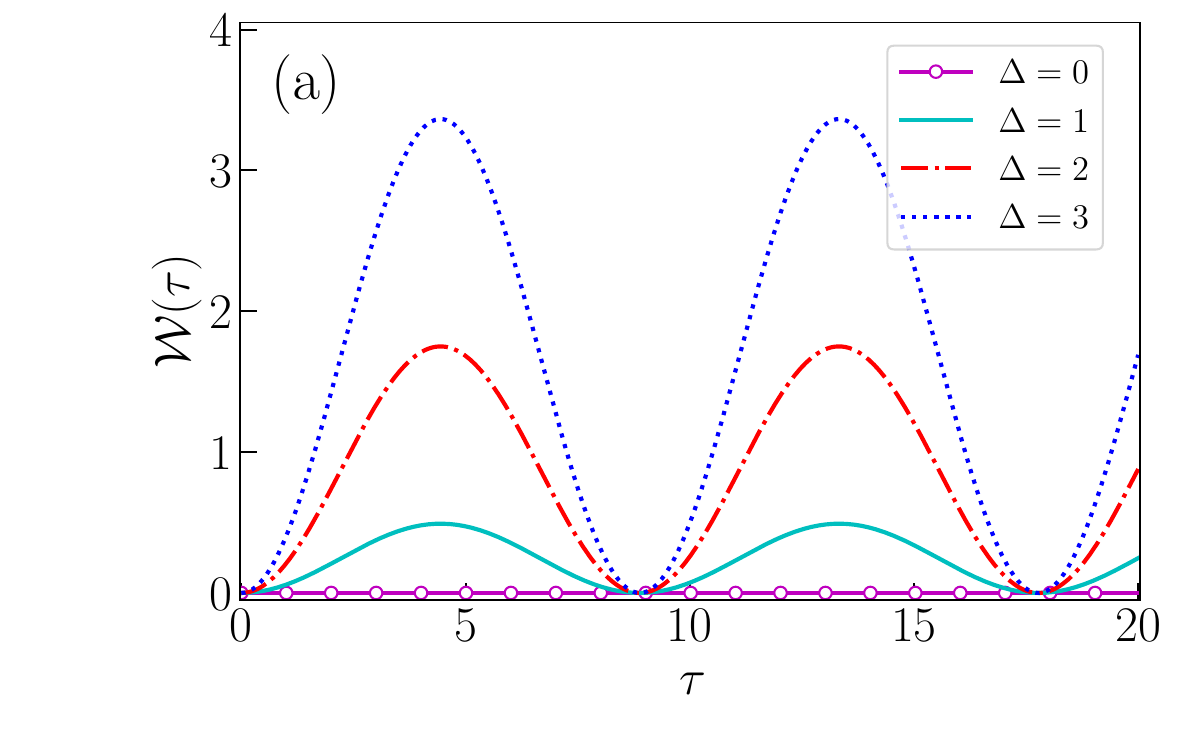}\label{}
        \includegraphics[width=0.3\textwidth, height=120px]{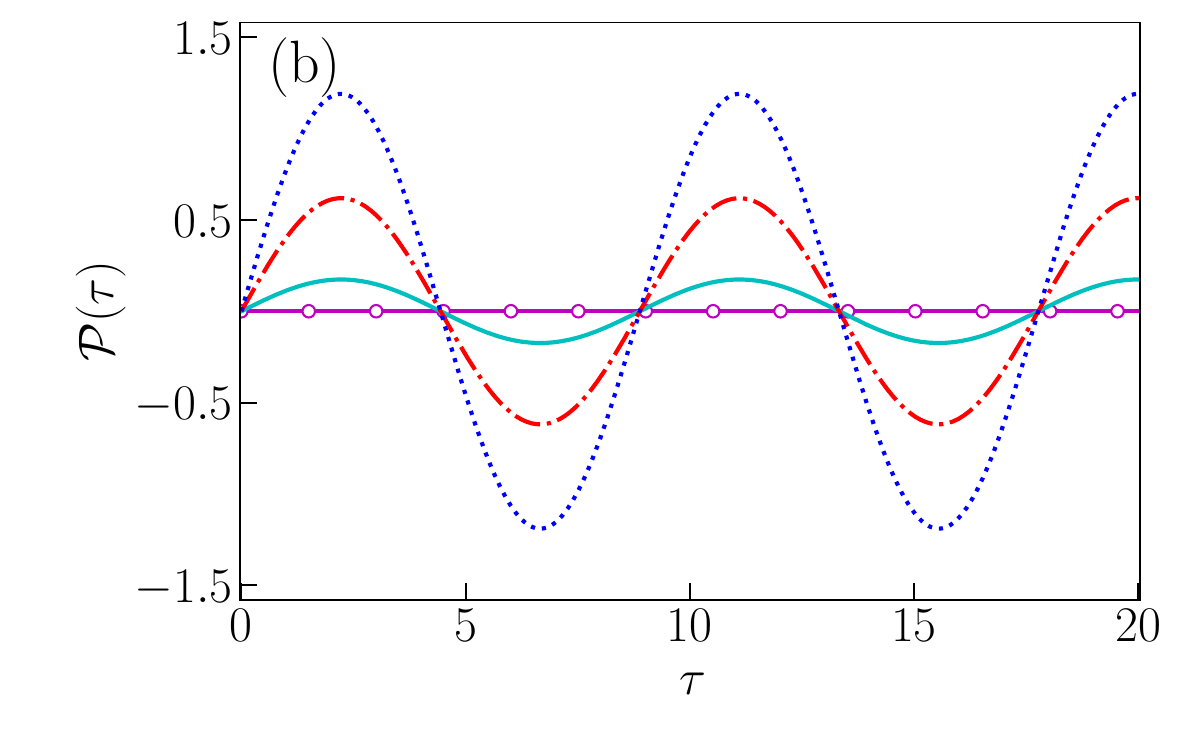}\label{}
         \includegraphics[width=0.3\textwidth, height=120px]{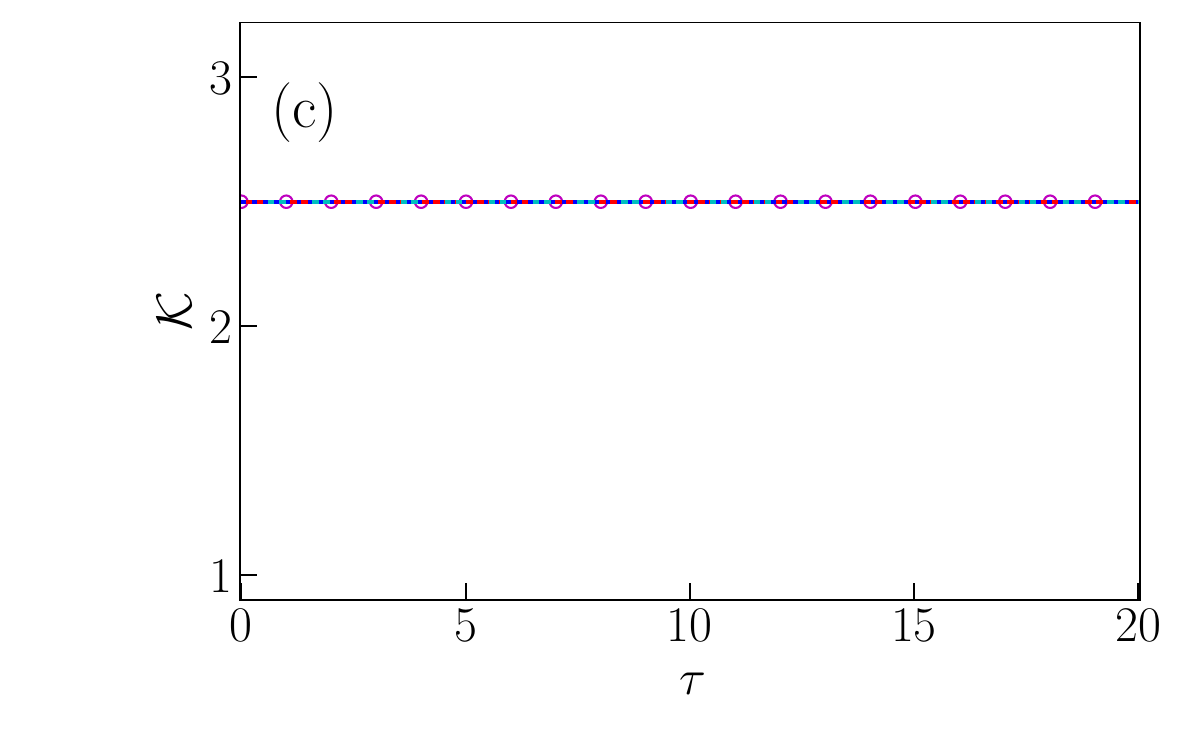}\label{}
        \end{center}
\caption{ Temporal evolution of (a) ergotropy, (b) power and (c) capacity of the hybrid quantum battery as a function of time at a few selected values of $\Delta$ with the fixed parameters~ $\theta=\pi/4$,~ $g_1=g_2=2$ and $\tilde{J}=\tilde{D}=\tilde{\Omega}=\tilde{T}=1$.} 
\label{del2}
\end{figure}
In addition to the effects of $\tilde{D}$, we systematically investigate the role of the exchange anisotropy parameter $\Delta$, which governs the nature of spin-spin interactions in the system. The exchange anisotropy parameter $\Delta$ plays a crucial role in governing the nature of spin interactions in the hybrid qubit–qutrit system. Physically, $\Delta$ controls the relative strength between the transverse (xy) and longitudinal (z) components of the Heisenberg interaction. As shown in Fig.(\ref{del1}), increasing $\Delta$ significantly enhances both the $l_1$-norm of coherence and negativity, indicating stronger quantum superposition and entanglement generation. This behavior arises because larger $\Delta$ amplifies the transverse coupling terms, which promote coherent transitions between spin states and facilitate population mixing.  The corresponding impact on the battery performance is illustrated in Fig. (\ref{del2}), where both ergotropy and charging power increase with $\Delta$, while the capacity remains unchanged. This demonstrates that exchange anisotropy acts as a key control parameter for optimizing energy storage and transfer in hybrid quantum batteries. We also observe that the peak ergotropy varies approximately linearly with the anisotropy parameter over the studied parameter range. The equation for fitting the peak ergotropy with respect to the parameter $\Delta$ is given by $\mathcal{W}(\Delta)_{max} \approx 1.137\Delta-0.227$

\begin{figure}[!h]
	\begin{center}
		\includegraphics[width=0.45\textwidth, height=145px]{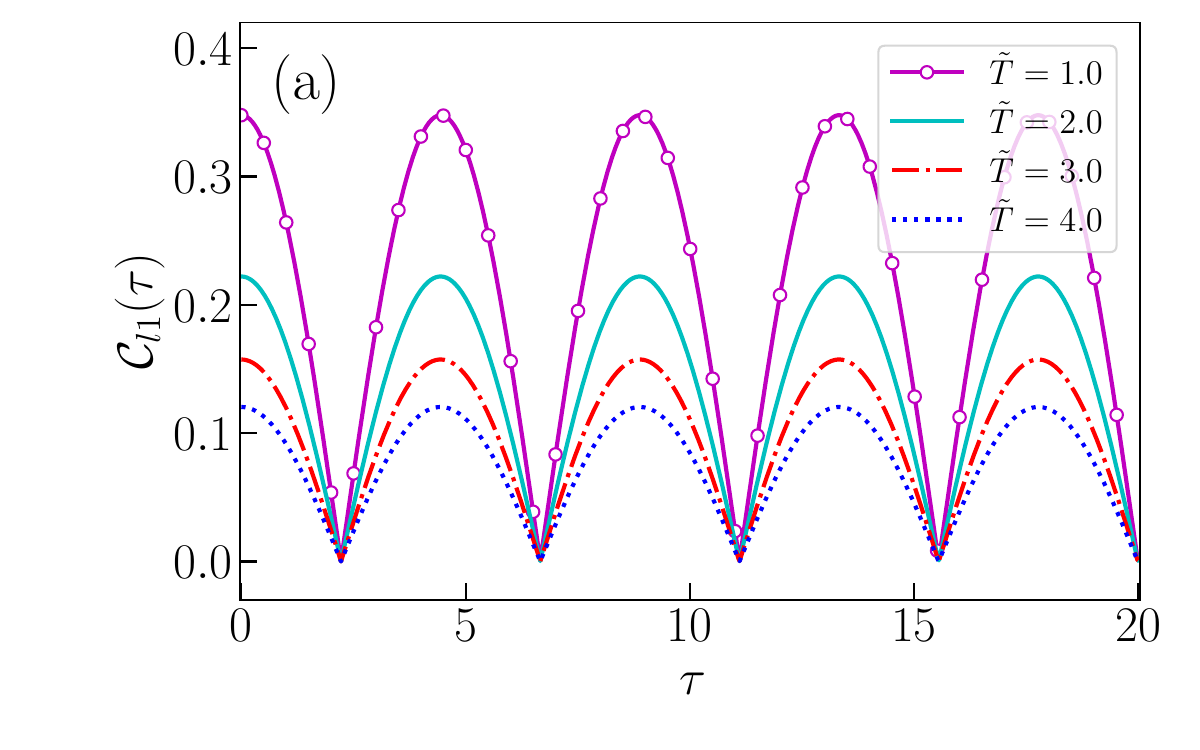} \label{3a}
	 	\includegraphics[width=0.45\textwidth, height=145px]{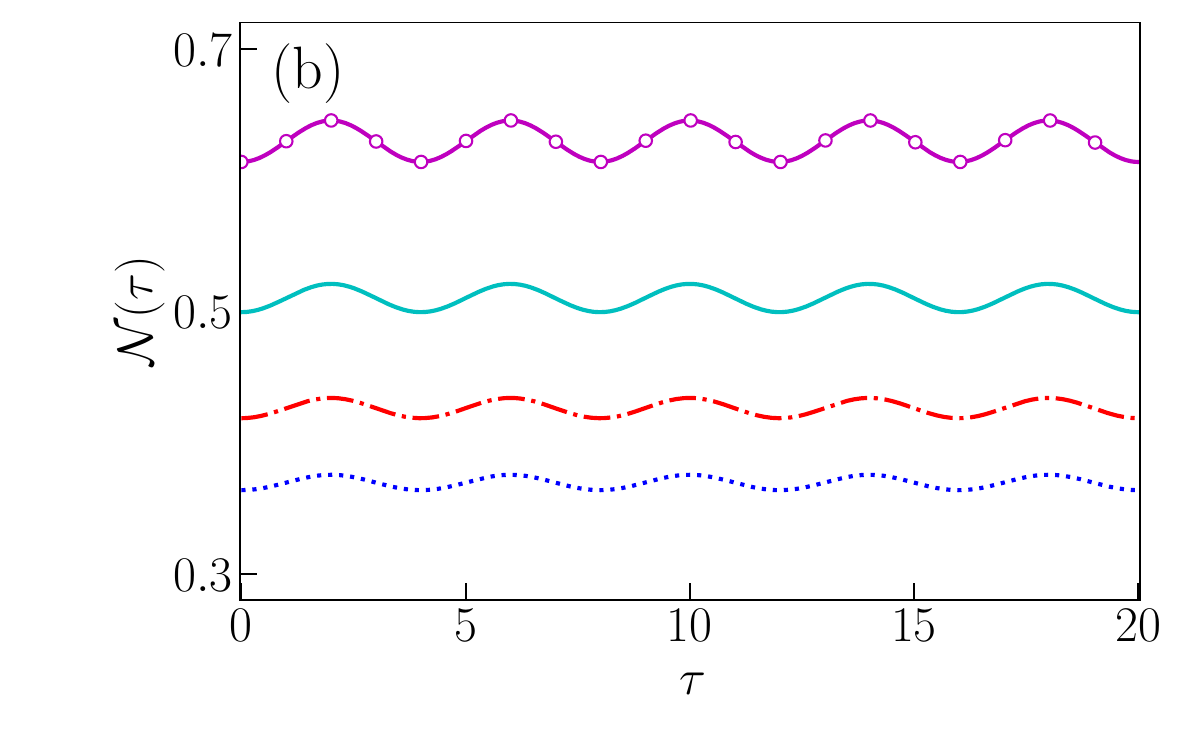}\label{3b}
        \end{center}
\caption{ Temporal evolution of (a) $l_1$-norm of coherence and (b) negativity of hybrid quantum battery as a function of time at a few selected values of $\tilde{T}$  with the fixed parameters  $\theta=\pi/4$, $g_{1}=g_{2}=2$ and $\tilde{J}=\tilde{D}=\tilde{\Omega}=B=1$ .} 
\label{fig3}
\end{figure}
In Fig. (\ref{fig3}), we report the temporal evolution of the  $l_1-$norm of coherence and negativity $\mathcal{N}(\rho)$ of the hybrid qubit–qutrit quantum battery for different operating temperatures $\tilde{T}=1,2,3, ~\text{and}~ 4$. As shown in Fig. (\ref{fig3}) a, the coherence exhibits periodic oscillations arising from the coherent unitary charging dynamics. Importantly, the oscillation amplitude decreases monotonically with increasing temperature $\tilde{T}$, indicating that thermal effects progressively suppress the quantum superposition available in the battery state. This reflects a loss of useful quantum resources under elevated thermal conditions. The right panel displays the corresponding dynamics of negativity, which quantifies entanglement between the qubit and qutrit subsystems. The persistence of oscillations confirms that entanglement is dynamically generated during the charging process. However, the overall magnitude of negativity is significantly reduced at higher temperatures, demonstrating that thermal fluctuations strongly limit the creation and preservation of nonclassical correlations.

These results establish temperature as a critical operational parameter for hybrid quantum batteries. Increasing thermal noise degrades both coherence and entanglement by washing out quantum correlations, revealing a clear competition between coherent driving (resource generation) and thermal mixing (resource degradation). The observed suppression of nonclassical resources provides a direct microscopic explanation for the deterioration of ergotropy and charging power at elevated temperatures. Consequently, Fig. (\ref{fig3}) highlights the importance of thermal engineering, noise suppression, and low-temperature operation for realizing efficient and scalable quantum battery devices.
\begin{figure}[!h]
\begin{center}
        \includegraphics[width=0.3\textwidth, height=120px]{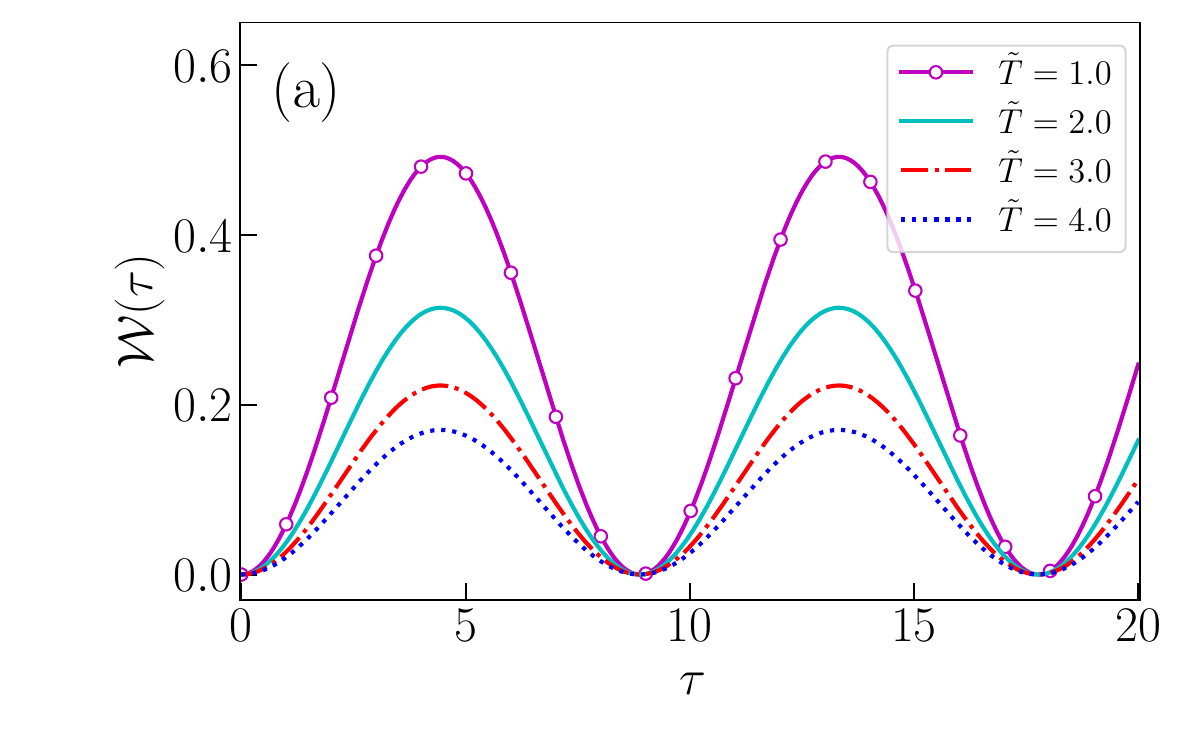}\label{4d}
        \includegraphics[width=0.3\textwidth, height=120px]{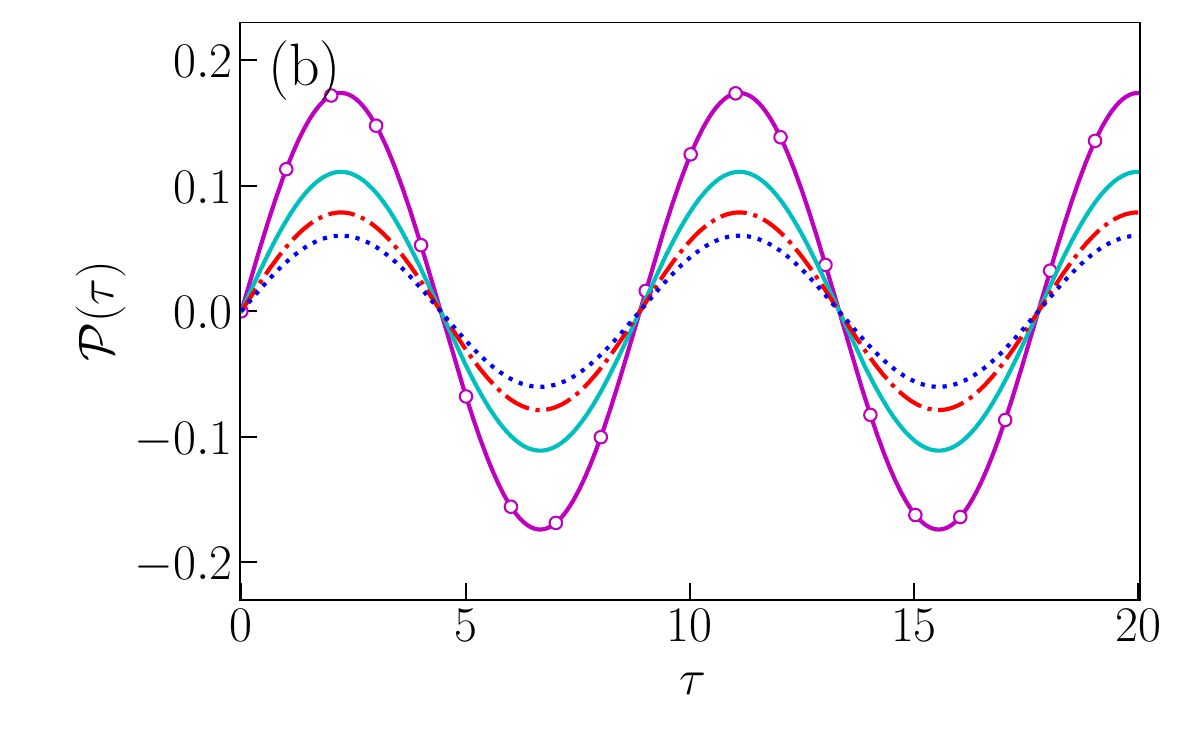}\label{4e}
         \includegraphics[width=0.3\textwidth, height=120px]{capT.pdf}\label{4f}
        \end{center}
\caption{ Temporal evolution of (a) ergotropy, (b) power and (c) capacity as a function of time at a few selected values of $\tilde{T}$ with the fixed parameters $\theta=\pi/4$, $g_{1}=g_{2}=2$ and $\tilde{J}=\tilde{D}=\tilde{\Omega}=B=1$ .} 
\label{fig4}
\end{figure}
Figure (\ref{fig4}) depicts the temporal evolution of the three key performance indicators of the qubit–qutrit quantum battery, namely ergotropy, charging power, and capacity, for different values of the system temperature $\tilde{T}$ (as mentioned in the above Figure). As shown in Fig. (\ref{fig4}) a, the ergotropy exhibits periodic oscillations in time due to the coherent unitary charging protocol, but its peak value decreases systematically with increasing temperature. Physically, this behavior originates from enhanced thermal mixing in the initial state at higher temperatures, which reduces population imbalance between the energy eigenstates. Consequently, the amount of useful extractable work stored in the battery is significantly reduced, indicating that elevated temperature directly limits the energy-storage capability of the device. The peak ergotropy shows a monotonic decay with temperature$\tilde{T}$, which is well described by an exponential-like suppression, consistent with thermal mixing effects. The fitting equation for this peak ergotropy fitting corresponding to the temperature is approximately given by $\mathcal{W}(\tilde{T})_{max}\approx -0.107~\tilde{T} +0.607$.

Figure (\ref{fig4}) b presents the corresponding charging power dynamics. The power retains its oscillatory behavior, confirming the persistence of coherent charging dynamics; however, the amplitude of power oscillations is progressively suppressed as temperature increases. This implies that thermal noise not only reduces the total stored energy but also degrades the effective charging rate, which is highly detrimental for practical device operation. In contrast, Fig. (\ref{fig4})c shows that the capacity $\mathcal{K}$ remains time independent and is largely insensitive to temperature. This behavior is expected, since the capacity depends solely on the spectral properties of the battery Hamiltonian, specifically the energy gap between the highest and lowest eigenstates, rather than on the state of the system. Therefore, temperature predominantly affects the state-dependent performance metrics (ergotropy and power), while leaving the intrinsic energy bandwidth of the device essentially unchanged.

Although the charging protocol remains coherent, increasing thermal fluctuations progressively suppress the quantum features of the state that are responsible for efficient work storage and rapid charging. This provides a direct operational connection to the results of Fig. (\ref{fig3}), where increasing temperature was shown to significantly reduce coherence and entanglement. Overall, Fig. (\ref{fig4}) highlights that efficient hybrid quantum batteries critically require stringent thermal control and noise-resilient architectures to preserve quantum advantage under realistic operating conditions.
\begin{figure}[!h]
\begin{center}
        \includegraphics[width=0.45\textwidth, height=145px]{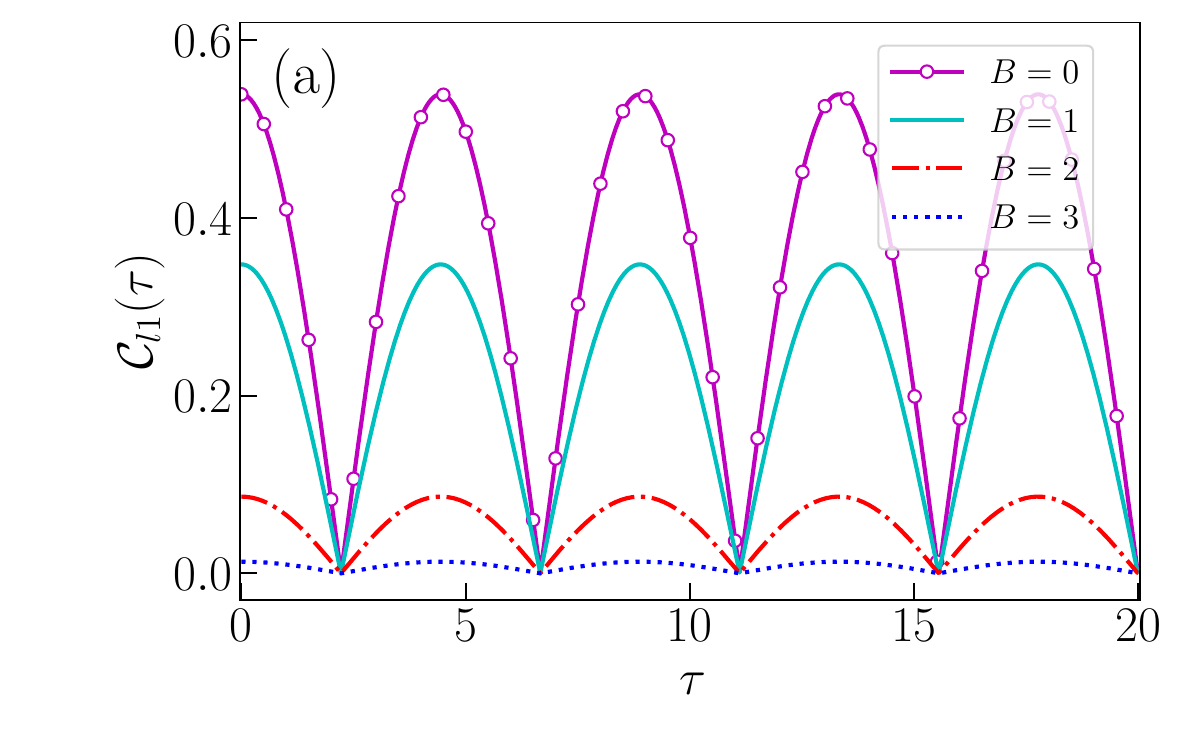}\label{}
        \includegraphics[width=0.45\textwidth, height=145px]{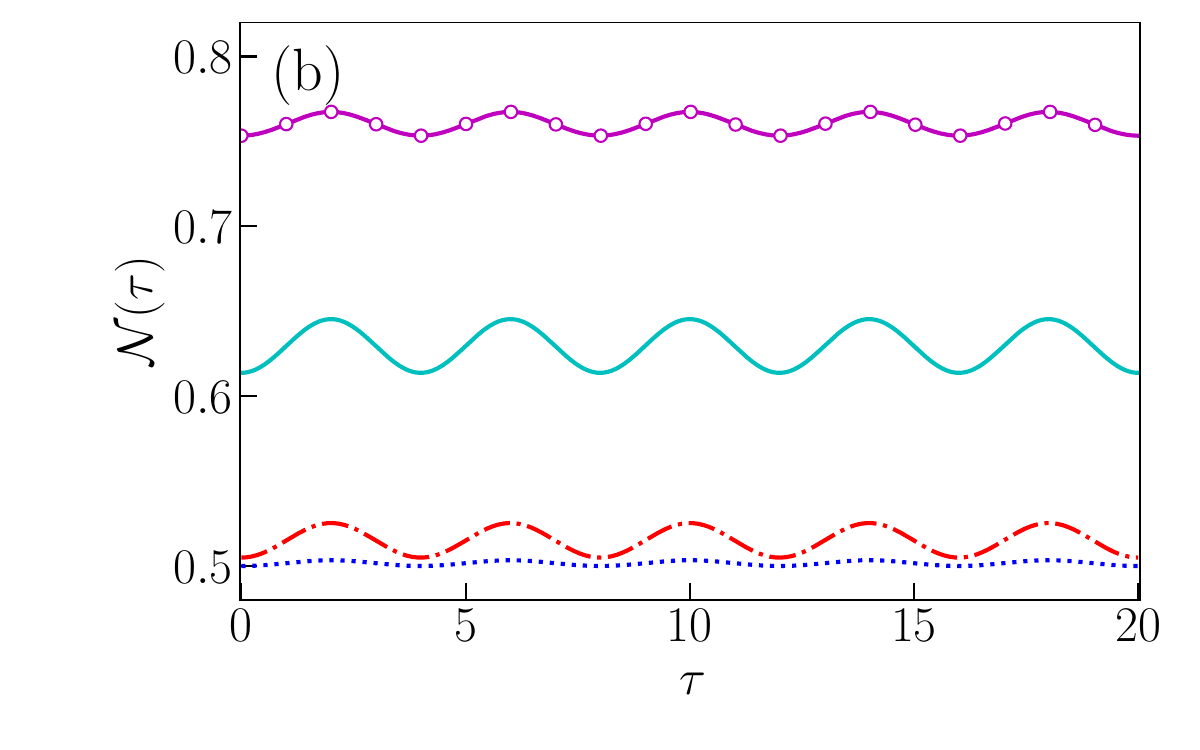}\label{}
        \end{center}
\caption{ Temporal evolution of (a) $l_1-$norm of coherence and (b) negativity of the hybrid quantum battery as a function of time $t$ for different values of $B$  with the parameters fixed $\theta=\pi/4$, $g_1=g_2=2$ and $\tilde{J}=\tilde{D}=\tilde{\Omega}=\tilde{T}=1$.} 
\label{fig7}
\end{figure}
Figure (\ref{fig7}) illustrates the time evolution of the $l_1-$norm of coherence and negativity of the qubit–qutrit quantum battery for several values of the applied magnetic field $B$. The oscillatory dynamics observed in both quantities originate from the coherent unitary evolution governed by the battery–charger Hamiltonian. The $l_1-$norm of coherence exhibits clear periodic oscillations, confirming that quantum superposition is dynamically generated and modulated during the charging process. Importantly, the amplitude and profile of these oscillations depend sensitively on the magnetic field strength. Increasing $B$ progressively modifies the energy level structure through Zeeman splitting, which alters the transition frequencies between eigenstates. As a result, the coherence dynamics become tunable via the external field, demonstrating that acts as an effective control parameter for engineering quantum resources in the battery. 

Similar to coherence, negativity displays oscillatory behavior, indicating periodic generation and degradation of entanglement during the charging cycle. The dependence on $B$ reveals that magnetic field tuning can either enhance or suppress entanglement generation depending on the regime, reflecting the competition between exchange interaction and Zeeman energy. Physically, stronger magnetic fields tend to favor product-like energy eigenstates, thereby reducing the ability of the interaction term to generate entanglement. Fig. (\ref{fig7}) highlights that the external magnetic field provides a practical and experimentally accessible control parameter for regulating nonclassical resources in hybrid quantum batteries. Since coherence and entanglement are directly linked to performance indicators such as ergotropy and charging power, magnetic-field engineering emerges as a viable strategy for controlling and optimizing the energetic and dynamical performance of the quantum battery.
\begin{figure}[!h]
\begin{center}
        \includegraphics[width=0.3\textwidth, height=120px]{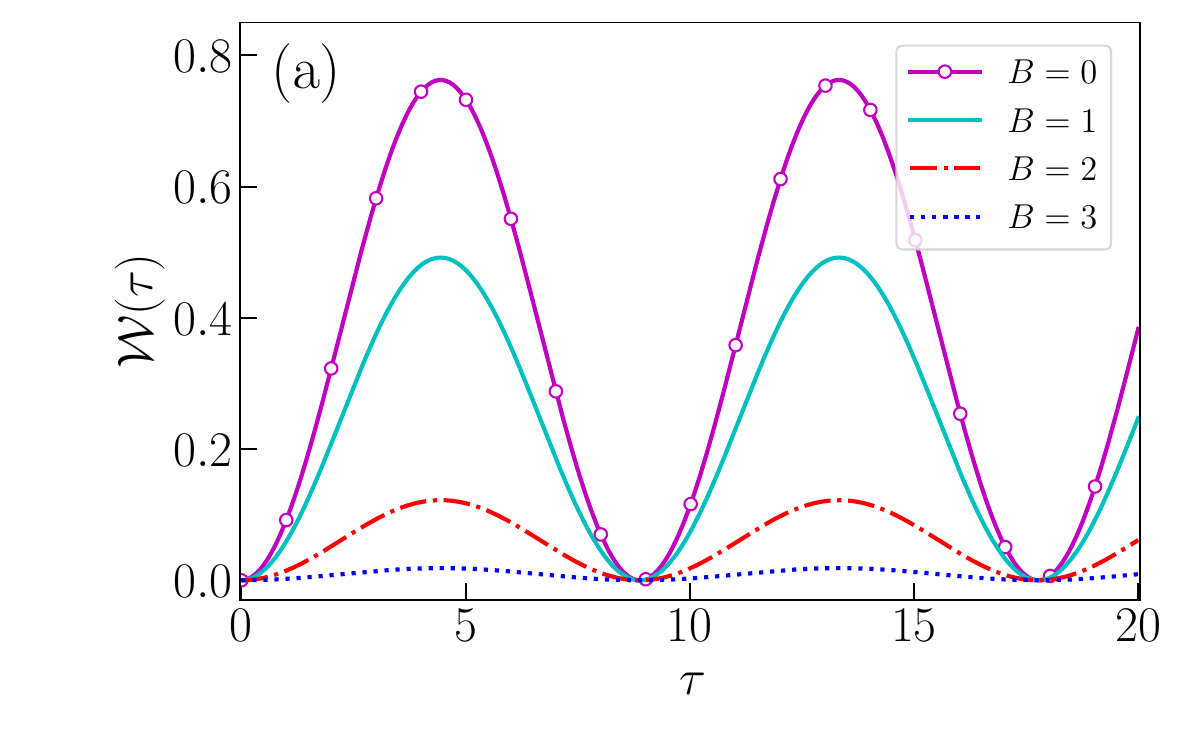}\label{}
        \includegraphics[width=0.3\textwidth, height=120px]{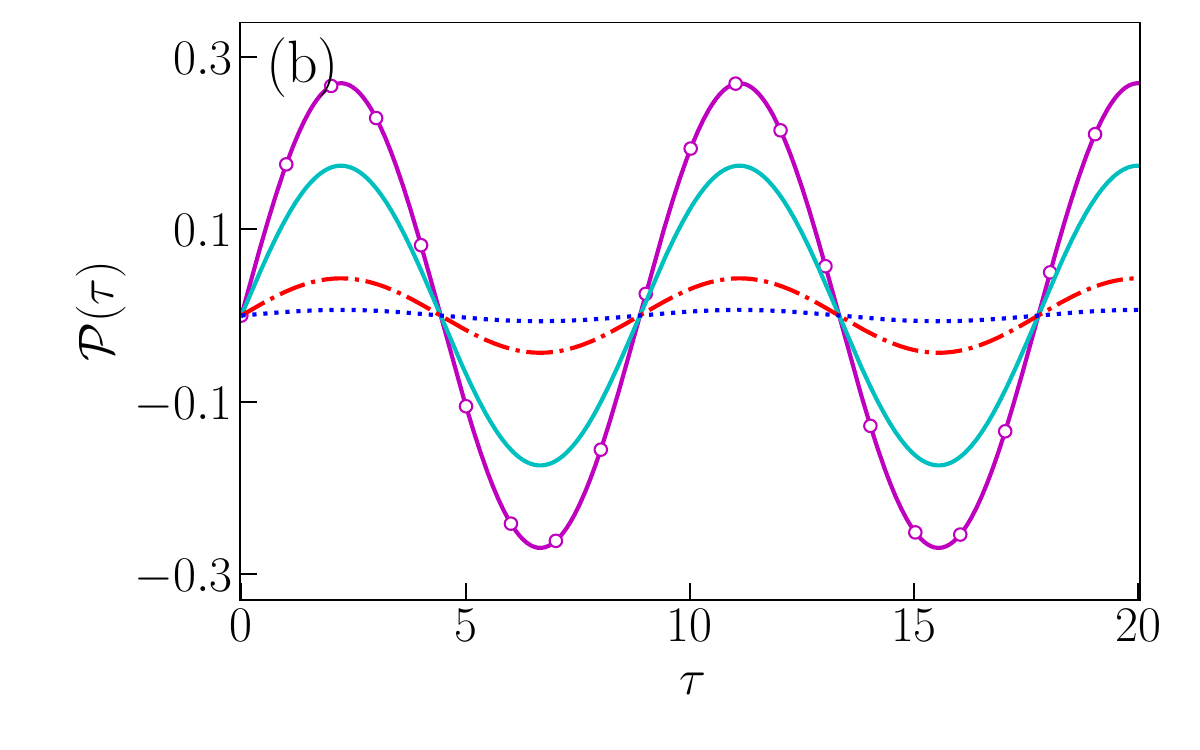}\label{}
         \includegraphics[width=0.3\textwidth, height=120px]{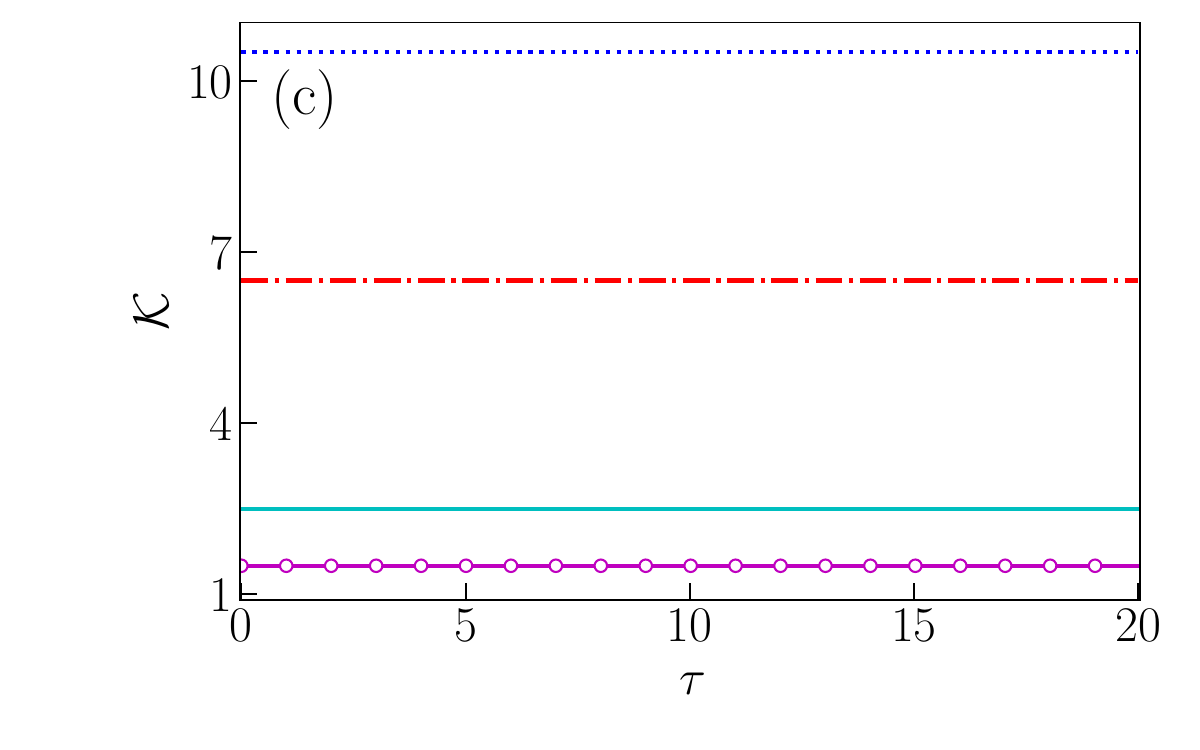}\label{}
        \end{center}
\caption{ Temporal evolution of (a) ergotropy, (b) power and (c) capacity of the hybrid quantum battery as a function of time at a few selected values of $B$ with the fixed parameters~ $\theta=\pi/4$,~ $g_1=g_2=2$ and $\tilde{J}=\tilde{D}=\tilde{\Omega}=\tilde{T}=1$.} 
\label{fig8}
\end{figure}

As shown in Fig. (\ref{fig8}) a, the ergotropy exhibits oscillatory dynamics due to coherent energy exchange between the battery and the charger. Both the amplitude and temporal profile of these oscillations are strongly dependent on the magnetic field strength, reflecting the field-induced modification of the energy spectrum through Zeeman splitting. Consequently, the population inversion responsible for extractable work becomes tunable with the magnetic field, enabling control over the stored usable energy. The dependence on the magnetic field $B$ is nonlinear and oscillatory, reflecting the interplay between exchange interaction and Zeeman splitting.The peak ergotropy fitting equation for the corresponding magnetic field is approximately given by $\mathcal{W}(B)_{max}\approx -0.250~B + 0.760 $. The corresponding charging power in Fig. (\ref{fig8}) b preserves the oscillatory structure but displays a systematic dependence on the magnetic field, with peak power varying with $B$. This indicates that the external field governs not only the amount of energy stored but also the rate at which it is delivered. In contrast, Fig. (\ref{fig8}) c shows that the capacity  $\mathcal{K}$ is time independent but shifts with the magnetic field, as expected from its dependence on the spectral width of the battery Hamiltonian. Together, these results demonstrate that magnetic-field tuning provides an effective means to control both the energetic and dynamical performance of the qubit–qutrit quantum battery.

Figures (\ref{fig7}) and (\ref{fig8}) unambiguously demonstrate that the external magnetic field acts as a unified control parameter governing both quantum resources and battery performance. The field-induced modulation of coherence and entanglement Fig.(\ref{fig7}) is directly reflected in the suppression of ergotropy and charging power Fig.(\ref{fig8}), establishing magnetic-field tuning as a powerful mechanism for engineering nonclassicality and optimizing the energetic and dynamical performance of hybrid qubit--qutrit quantum batteries. 

It is more important to highlight the direct operational link between nonclassicality and battery performance. We find that quantum coherence and entanglement are positively correlated with ergotropy and charging power across all investigated parameter regimes. Moreover, the analysis provides a practical optimization strategy, indicating that enhanced anisotropic interactions together with reduced thermal fluctuations maximize both quantum resources and the energetic and dynamical performance of the quantum battery. These results establish nonclassicality as a useful resource for the design of high-performance hybrid quantum batteries.

    

\section{ Nickel-Radical quantum battery}

Nickel-radical molecular complex   $(\text{Et}_3\text{NH})[\text{Ni(hfac)}_2\text{L}]$ provides a natural realization of a qubit–qutrit system in the absence of single-ion anisotropy \cite{Nickel2,Nickel3,Nickel}. In the following, we analyze the nonlocal correlations and battery performance metrics of this nickel–radical quantum battery. Figure (\ref{fig10}) presents the time evolution of quantum resources (coherence $C_{l_1}(\tau)$ and  negativity $\mathcal{N}(\tau)$) and performance indicators (ergotropy $\mathcal{W}(\tau)$, power $\mathcal{P}(\tau)$ and capacity $\mathcal{K}$) for the nickel–radical molecular quantum battery at different temperatures.

The quantum coherence and entanglement exhibit clear periodic oscillations, confirming that nonclassical correlations are dynamically generated and sustained even in the realistic molecular system. Although temperature affects the amplitude slightly, these quantum features persist, indicating robustness under experimental conditions. The ergotropy and power also show oscillatory behavior, reflecting cyclic energy exchange during the charging process, while the capacity remains constant over time, as it depends only on the system Hamiltonian. Importantly, the large magnitude of ergotropy demonstrates that the molecular system can store and deliver significant extractable work. Furthermore, the figure confirms that the proposed qubit–qutrit quantum battery retains both quantum resources and efficient energy-storage performance under realistic, experimentally relevant conditions.
\begin{figure}[!h]

        \includegraphics[width=0.45\textwidth, height=145px]{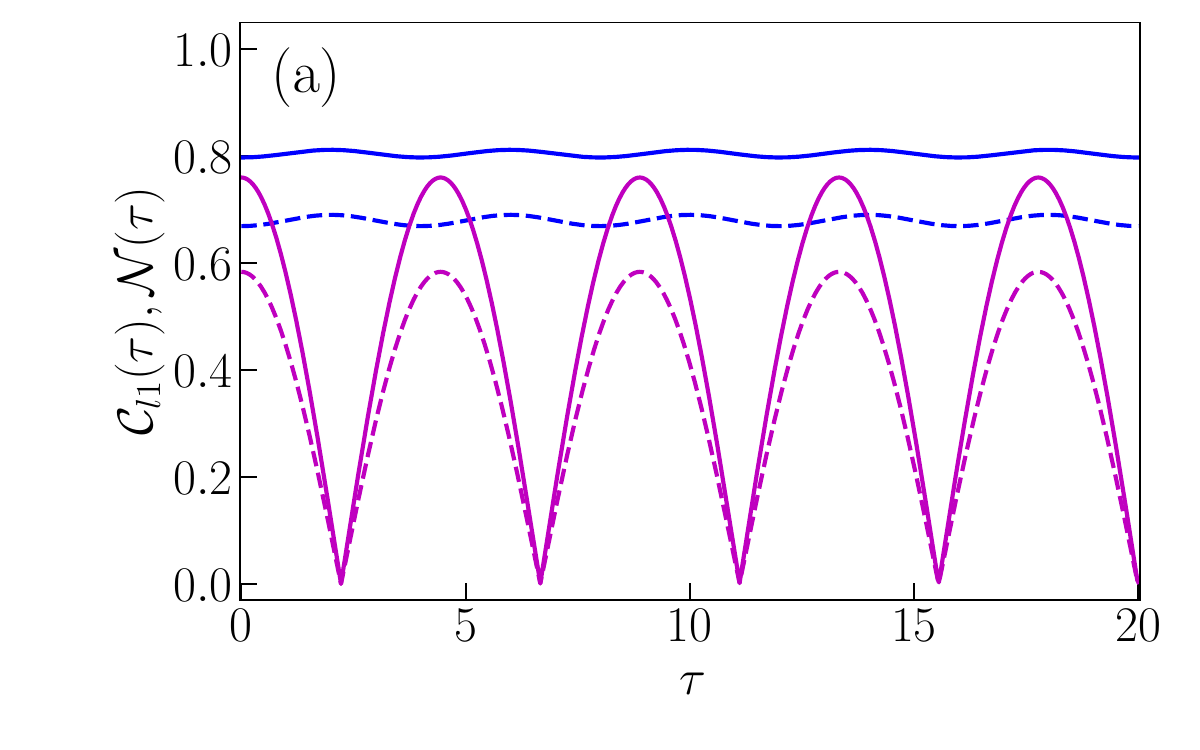}\label{}
         \includegraphics[width=0.45\textwidth, height=145px]{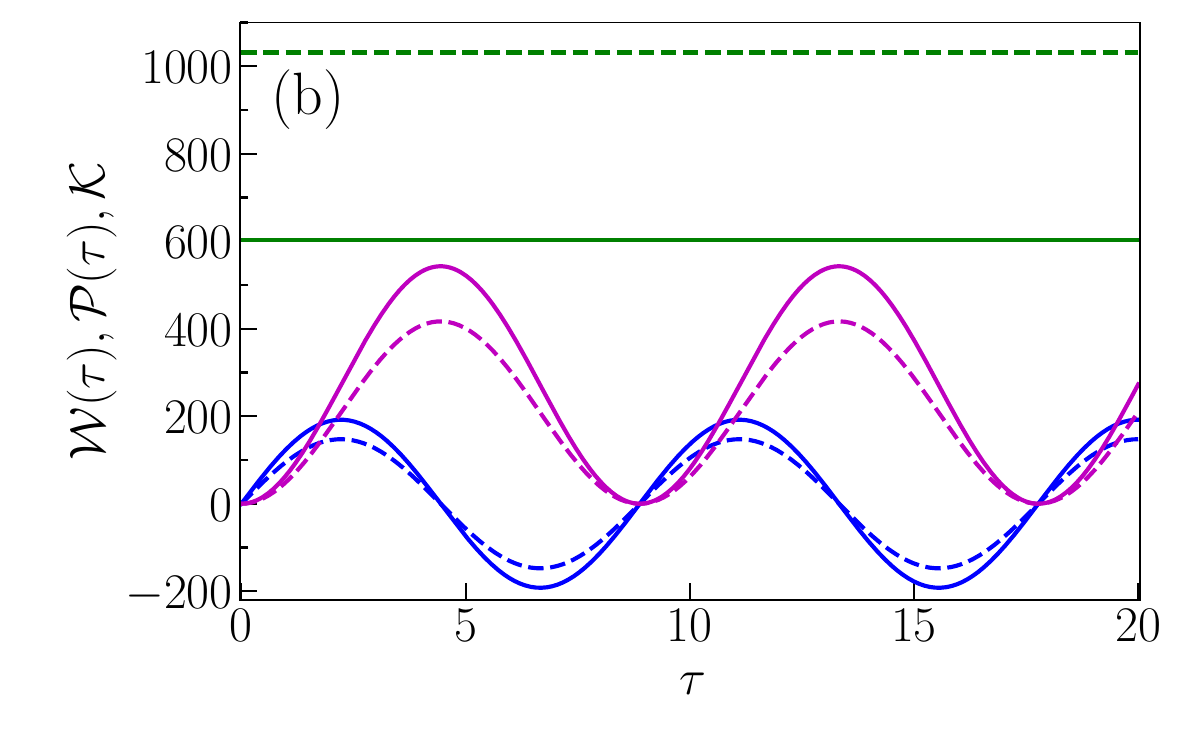}\label{}
\caption{(a) Temporal evolution of quantumness measures such as $l_1-$norm of coherence (pink line) and negativity (blue line)  and. (b) The ergotropy (pink line) , power (blue line) and capacity (green line) of hybrid quantum battery as a function of time at a few selected values of $B$ ( solid line for $B=200\tilde{T}$ and dashed line for $B=300\tilde{T}$) for the nickel-radical molecular compound  with the fixed parameters $J/k_B = 505K $, $\tilde{D}=0$, $g_1=2.005$, $g_2=2.275$ and $\tilde{\Omega}=1$.}
\label{fig9}
\end{figure}

\begin{figure}[!h]
\begin{center}
        \includegraphics[width=0.45\textwidth, height=145px]{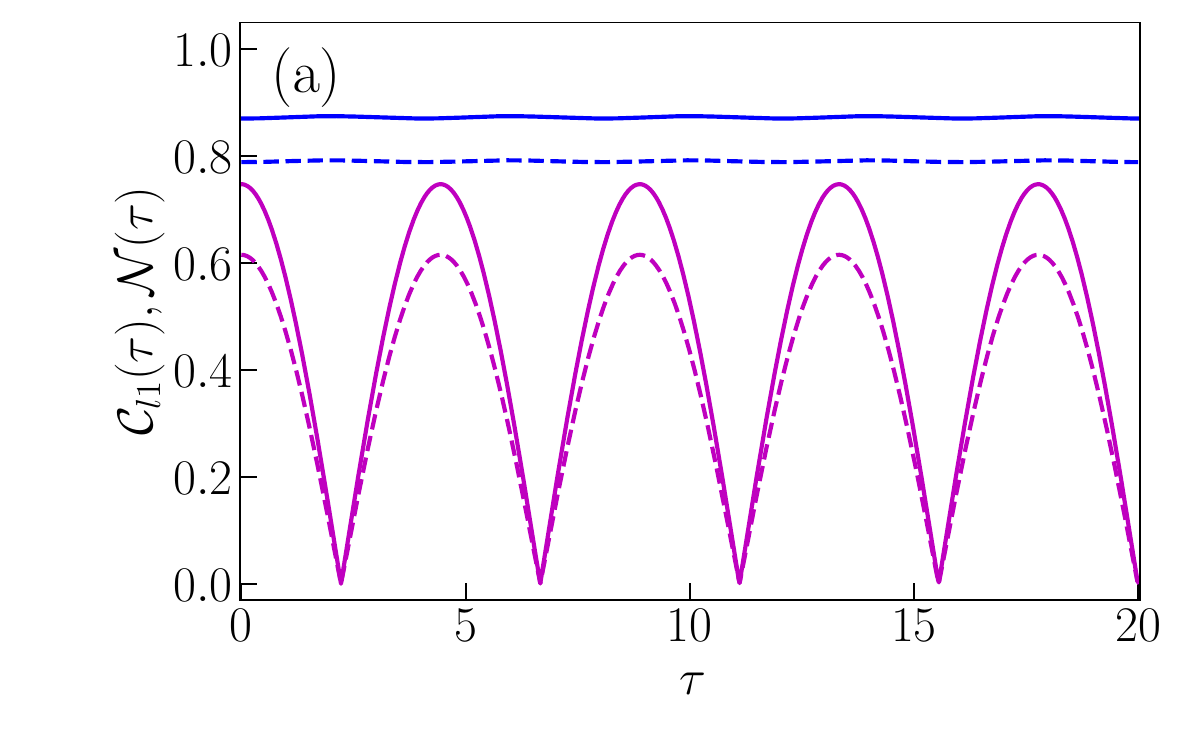}\label{}
         \includegraphics[width=0.45\textwidth, height=145px]{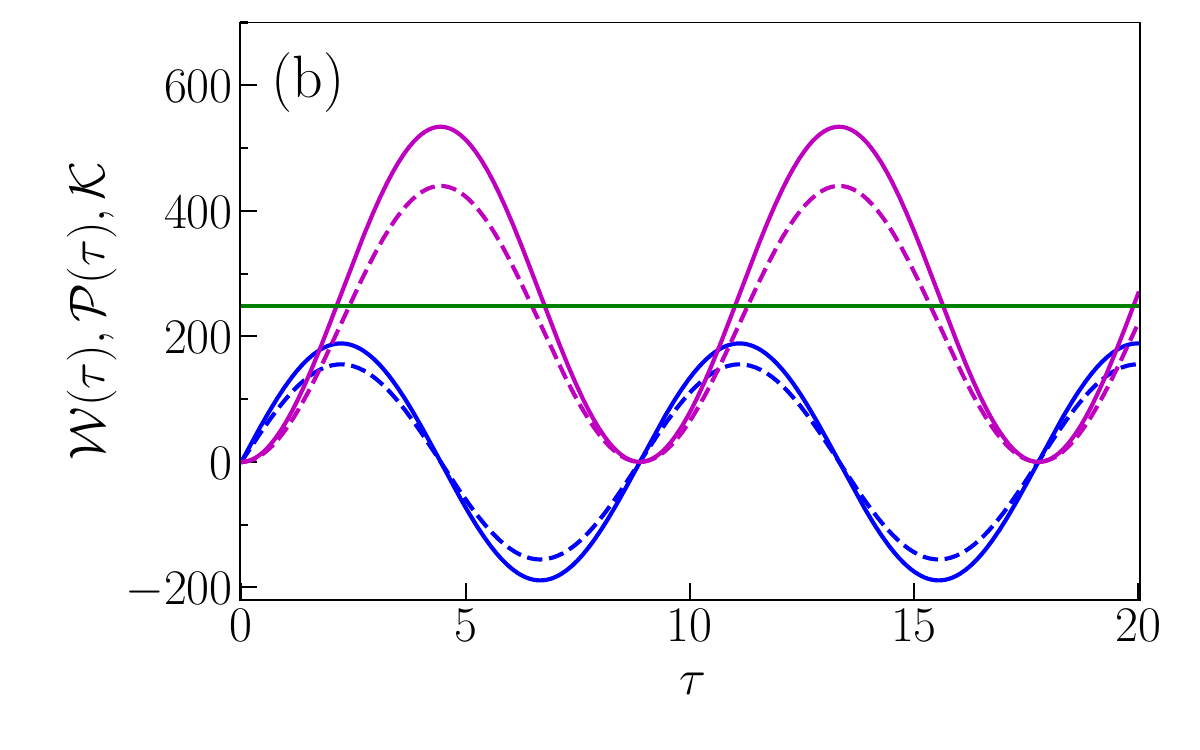}\label{}
        \end{center}
\caption{(a) Temporal evolution of quantumness measures such as $l_1-$norm of coherence (pink line) and negativity (blue line) and. (b) The ergotropy (pink line), power (blue line) and capacity (green line) of hybrid quantum battery as a function of time at a few selected values of $\tilde{T}$ ( solid line for $\tilde{T}=300K$ and dashed line for $\tilde{T}=400K$) for the nickel-radical molecular compound with the fixed parameters set $J/k_B = 505 $, $\tilde{D}=0$, $g_1=2.005$, $g_2=2.275$ and $\tilde{\Omega}=1$.}
\label{fig10}
\end{figure}

Figure (\ref{fig9}) illustrates the dependence of quantum resources (coherence and entanglement) and performance indicators (ergotropy and power) on the magnetic field for the nickel–radical molecular quantum battery. The results show that both quantum coherence and entanglement decrease with increasing magnetic field, eventually vanishing beyond a critical field due to spin alignment induced by the Zeeman interaction. Similarly, ergotropy and power are strongly influenced by the magnetic field, exhibiting reduced amplitudes at higher field strengths. This behavior indicates that strong magnetic fields suppress nonclassical correlations and limit the energy-storage capability of the quantum battery. Therefore, an optimal low-to-moderate magnetic-field regime is essential for maintaining both quantum advantages and efficient battery performance.

\section{Experimental relavance}
The proposed qubit–qutrit quantum battery model is not merely of theoretical interest but admits a realistic physical implementation in molecular magnetic platforms, particularly in the nickel–radical molecular complex   $(\text{Et}_3\text{NH})[\text{Ni(hfac)}_2\text{L}]$. From a microscopic perspective, this compound naturally realizes a mixed spin-(1/2, 1) Heisenberg dimer, where a spin-1/2 nitronyl–nitroxide radical is exchange coupled to a spin-1 $\mathrm{Ni}^{2+}$ ion. This architecture maps directly onto the qubit–qutrit battery Hamiltonian employed in our model. Importantly, this molecular system is characterized by a large antiferromagnetic exchange coupling $(J/k_B=505)$  and distinct Landé $g-$factors, which together enable robust quantum correlations and strong controllability under external magnetic fields \cite{Nickel3}. The quantum resource content of the nickel–radical compound is examined in detail, revealing that both entanglement and coherence persist even at room temperature \cite{Nickel2,Nickel}. Moreover, the model parameters are carefully selected to align with experimentally reported values of the nickel–radical complex, ensuring that the dynamics of ergotropy, charging power, coherence, and entanglement remain well-defined and controllable within experimentally accessible magnetic field regimes. This establishes that the predicted battery performance is not an abstract idealization but corresponds to physically achievable operating regimes.

The charging protocol considered in our model can be implemented experimentally through resonant microwave or radio-frequency magnetic driving using established electron spin resonance (ESR) techniques. An oscillatory magnetic field applied along the quantization axis realizes the effective charger Hamiltonian, where the charging strength $\tilde{\Omega}$ corresponds to the amplitude of the applied field. By tuning the microwave pulse intensity and duration, one can directly control the charging rate and energy transfer between the qubit and qutrit subsystems. This provides a practical and experimentally accessible mechanism to simulate the cyclic charging dynamics predicted by the theoretical model.

The energetic performance of the molecular quantum battery can be inferred from population redistribution among spin eigenstates, which is measurable via time-resolved ESR spectroscopy and magnetization dynamics. The extractable work (ergotropy) corresponds to the degree of population inversion generated during the driving cycle, while charging power is reflected in the rate of change of these populations. Quantum coherence can be probed using Ramsey-type interferometric protocols, and signatures of entanglement may be reconstructed through density-matrix tomography techniques adapted for spin dimers. Importantly, previous experimental studies on nickel–radical molecular magnets report observable quantum correlations and long-lived spin coherence near room temperature, indicating that the predicted oscillatory battery dynamics occur within experimentally accessible decoherence windows.

Using experimentally realistic parameters, our simulations demonstrate persistent coherence and entanglement at temperatures $\tilde{T}$ = 300–400 K \cite{Nickel2,Nickel}, establishing that the battery operates beyond cryogenic conditions. For the nickel--radical molecular realization, we again choose $E_0=J$, with $J/k_B=505\,{\rm K}$. Hence, $J\simeq 43.5\,{\rm meV}$ and the corresponding physical time unit is $\hbar/J\simeq 15.1\,{\rm fs}$. A physical driving frequency is converted according to $\tilde{\Omega}=\hbar\Omega/J$; for example, $\Omega/(2\pi)=1\,{\rm GHz}$ corresponds to $\tilde{\Omega}\simeq9.5\times10^{-5}$. Although thermal fluctuations slightly suppress oscillation amplitudes, the nonclassical features and performance indicators remain well defined, confirming the robustness of the device. This stability originates from the strong exchange coupling and chemical protection of the molecular spin architecture. Molecular magnetic systems further offer advantages including chemical tunability, nanoscale scalability, and precise external field control, making them promising candidates for the realization of hybrid quantum energy-storage devices.

The nickel–radical molecular complex provides a concrete pathway toward experimental realization of qubit–qutrit quantum batteries. The direct mapping between the theoretical Hamiltonian and an existing molecular magnet, combined with established ESR control and measurement techniques, demonstrates that the proposed battery model lies within current experimental capabilities. These results open a viable route toward the development of room-temperature molecular quantum energy storage architectures. 
Realistic molecular systems are subject to spin–phonon coupling, thermal fluctuations, and material imperfections, which can reduce coherence and charging performance through decoherence processes. However, owing to the strong exchange coupling and experimentally observed robustness of quantum correlations in nickel–radical molecular complexes, the main features of the proposed quantum battery are expected to remain observable under realistic operating conditions.

\section{Conclusions}
\label{cncl}
In this work, we have investigated a hybrid qubit--qutrit quantum battery based on a mixed spin-$1/2$--spin-$1$ system interacting via
anisotropic Heisenberg exchange coupling and driven by an external magnetic field. The nonclassical properties were characterized using the $l_1$-norm of coherence and negativity, while the energy-storage performance was evaluated in terms of ergotropy, power, and capacity. Our results demonstrate the presence of strong nonclassical correlations and reveal a close correspondence between the evolution of these quantum resources and battery-performance indicators such as ergotropy and charging power. Both ergotropy and power exhibit oscillatory charging--discharging dynamics, whereas the capacity remains time independent and is determined by the battery Hamiltonian. The effects of the system parameters and magnetic field further demonstrate a strong interplay between quantum resources, ergotropy, and charging power.

Furthermore, we established a direct connection between the proposed model and the experimentally realizable nickel--radical molecular
complex $(\mathrm{Et}_3\mathrm{NH})[\mathrm{Ni(hfac)}_2\mathrm{L}]$. Using experimentally relevant parameters, we find that quantum
coherence, entanglement, and appreciable energy-storage performance persist even at room temperature, demonstrating the robustness of the proposed system under realistic conditions. These results provide a promising pathway toward the realization of hybrid qubit--qutrit quantum batteries in solid-state molecular platforms. Future studies may further investigate decoherence and spin--phonon effects in realistic molecular quantum batteries.

\onecolumngrid
\appendix
\section{Thermal state of hybrid Quantum Battery}
\label{A}
Using the definition of thermal state Eq. (3), the density matrix of the qubit-qutrit system can be computed as
\begin{align}
    \varrho(0,\tilde{T})=\frac{1}{\mathcal{Z}}
    \begin{pmatrix}
    \varrho_{11}^0&0&0&0&0&0\\
    0& \varrho_{22}^0&0& \varrho_{24}^0&0&0\\
    0&0& \varrho_{33}^0&0& \varrho_{35}^0&0\\
    0& \varrho_{42}^0&0& \varrho_{44}^0&0&0\\
    0&0& \varrho_{53}^0&0& \varrho_{55}^0&0\\
    0&0&0&0&0& \varrho_{66}^0
    \end{pmatrix}, 
\end{align}
and the matrix elements are\\[0.5cm]
\begin{align}
  \varrho_{11}^0&=e^{\frac{-\beta \alpha_-}{2}}, \;\;\;\;\;
  \varrho_{66}^0=e^{\frac{-\beta \alpha_+ }{2}}, \nonumber \\  
 \varrho_{22}^0&=e^{\frac{\beta \Lambda_+}{4}}\left[\cosh \left(\frac{\beta \eta _-}{4}\right)-\frac{\sinh \left(\frac{\beta \eta _-}{4}\right) \gamma_-}{\eta _-}\right], \nonumber\\
\varrho_{33}^0&=e^{\frac{\beta \Lambda_-}{4} }\left[\cosh \left(\frac{\beta \eta _+}{4}\right)+\frac{\sinh \left(\frac{\beta \eta _+}{4}\right) \gamma_+}{\eta _+}\right], \notag \\
\varrho_{44}^0&=e^{\frac{\beta \Lambda_+}{4}}\left[\cosh \left(\frac{\beta \eta _-}{4}\right)+\frac{\sinh \left(\frac{\beta \eta _-}{4}\right) \gamma_-}{\eta _-}\right],\notag \\
\varrho_{55}^0&=e^{\frac{ \beta \Lambda_-}{4} }\left[\cosh \left(\frac{\beta \eta _+}{4}\right)-\frac{\sinh \left(\frac{\beta \eta _+}{4}\right) \gamma_+}{\eta _+}\right],\notag \\
\varrho_{24}^0&=\varrho_{42}^0=\frac{1}{\eta _-}\left({-\sqrt{8}\left(\tilde{J}\Delta \right) e^{\frac{ \beta \Lambda_+}{4} }} \sinh \left(\frac{\beta \eta _-}{4}\right) \right), \notag \\
\varrho_{35}^0&=\varrho_{53}^0=\frac{1}{\eta _+}\left({-\sqrt{8}\left( \tilde{J}\Delta \right)  e^{\frac{\beta \Lambda_-}{4}}\sinh \left(\frac{\beta \eta _+}{4}\right)}\right). \notag
\end{align}\\
and the partition function  $\mathcal{Z}$ of the system is \\
\begin{align}
   \mathcal{Z}=2\left[e^{\frac{-\beta(\tilde{J}+2\tilde{D})}{2}}\cosh\left({\frac{\beta(\textcolor{black}{\tilde{h}_1+2\tilde{h}_2)}}{2}}\right)+e^{{\frac{\beta(\tilde{J}-2\tilde{D})}{4}}}\left[e^{\frac{\beta \tilde{h}_2}{2}}\cosh\left(\frac{\beta \eta_-}{4}\right)+e^{\frac{-\beta \tilde{h}_2}{2}}\cosh\left(\frac{\beta \eta_+}{4}\right)\right]\right]
\end{align}
with $\eta_\pm=\sqrt{[\tilde{J}-2\tilde{D}\pm2(\tilde{h}_1-\tilde{h}_2)]^2} + 8(\tilde{J}\Delta)^2$, $\alpha_\pm=\tilde{J}+2\tilde{D}\pm(\tilde{h}_1 +2\tilde{h}_2)$, $\gamma_\pm=\tilde{J}-2\tilde{D}\pm2(\tilde{h}_1 -\tilde{h}_2)$ and $\Lambda_{\pm}=-2\tilde{D}\pm 2 \tilde{h}_2+\tilde{J}$.

\section{Time evolved state hybrid quantum battery}
\label{evoleved}
After applying the charging protocol, the time-evolved density matrix $\varrho(\tau,\tilde{T})$ of the QB is calculated as
\begin{align} 
\varrho(\tau,\tilde{T})=\frac{1}{\mathcal{Z}}
    \begin{pmatrix}
        \varrho^{\tau}_{11}&0&0&0&0&0\\
        0&\varrho^{\tau}_{22}&0&\varrho^{\tau}_{24}&0&0\\
        0&0&\varrho^{\tau}_{33}&0&\varrho^{\tau}_{35}&0\\
        0&\varrho^{\tau}_{42}&0&\varrho^{\tau}_{44}&0&0\\
        0&0&\varrho^{\tau}_{53}&0&\varrho^{\tau}_{55}&0\\
        0&0&0&0&0&\varrho^{\tau}_{66}\\
    \end{pmatrix}
\end{align}
where the non-zero matrix elements are
\begin{align}
\varrho^{\tau}_{11,66}&= e^{-\frac{\beta}{2}\left(2\tilde{D}\mp\tilde{h}_1\mp2\tilde{h}_2+\tilde{J}\right)},\notag \\
\varrho^{\tau}_{33,55}&= e^{\frac{\beta\Lambda_-}{4}}\left[\cosh\left(\frac{\beta\eta_+}{4}\right)\pm\frac{\sinh\left(\frac{\beta\eta_+}{4}\right)\gamma_+}{\eta_+}\right],\notag \\
\varrho^{\tau}_{42,24}&= -\frac{2\sqrt{2}\tilde{J}\Delta}{\eta_-}\sinh\left(\frac{\beta\eta_-}{4}\right)\exp\left[\pm i\tilde{\Omega}\tau(\cos\theta-2\sin\theta)+\frac{\beta\Lambda_+}{4}\pm2i\tilde{\Omega}\sin\theta\,\tau\right],\notag \\
\varrho^{\tau}_{53,35}&= -\frac{2\sqrt{2}\tilde{J}\Delta}{\eta_+}\sinh\left(\frac{\beta\eta_+}{4}\right)\exp\left[\mp i\tilde{\Omega}\tau(2\sin\theta-\cos\theta)+\frac{\beta\Lambda_-}{4}\pm2i\tilde{\Omega}\sin\theta\,\tau\right],\notag \\
\varrho^{\tau}_{22}&= e^{\frac{\beta\Lambda_+}{4}}\left[\cosh\left(\frac{\beta\eta_-}{4}\right)-\frac{\alpha_-}{\eta_-}\sinh\left(\frac{\beta\eta_-}{4}\right)\right],\notag \\
\varrho^{\tau}_{44}&= e^{\frac{\beta\Lambda_+}{4}}\left[\cosh\left(\frac{\beta\eta_-}{4}\right)+\frac{\gamma_-}{\eta_-}\sinh\left(\frac{\beta\eta_-}{4}\right)\right].\notag
\end{align}
\section{Time evolution of Quantum battery under environmental noise}
The battery's dynamics under an open quantum system is not an unitary evolution and its governed by the Lindblad master equation, which is given by\\
\begin{align}
    \dot{\varrho}(\tau)= \frac{-i}{\hbar}[\mathcal{H},\varrho(\tau)]+\gamma\mathcal{L}[\varrho(\tau)]
\end{align}
where $\gamma$ is the dissipation strength, $\mathcal{H}$ is the total Hamiltonian of the system including the battery, charger and the interaction between them. $\mathcal{L}(\varrho)$ is known as the Lindblad operator which is given by\\
\begin{align}
    \mathcal{L}[\varrho(\tau)]=\sum_i[2L_i\varrho(\tau) L_i^{\dagger}-\{L_i^{\dagger} L_i , \varrho(\tau) \}]
\end{align}
where $L_i=\hat{\sigma}^{z}_{i}\otimes \mathbb{I}$, $\hat{\sigma^z_i}$~is the Pauli-Z matrix acting on the $i^{\rm th}$ qubit.

Thus with the $\varrho(\tau)$ obtained by solving the master equation is used to compute the $l_1$ norm of coherence and ergotropy  which captures the quantumness and performance of the quantum battery under environmental conditions.
\begin{figure}[!h]
    \centering
    \includegraphics[width=0.40\textwidth,height=125px]{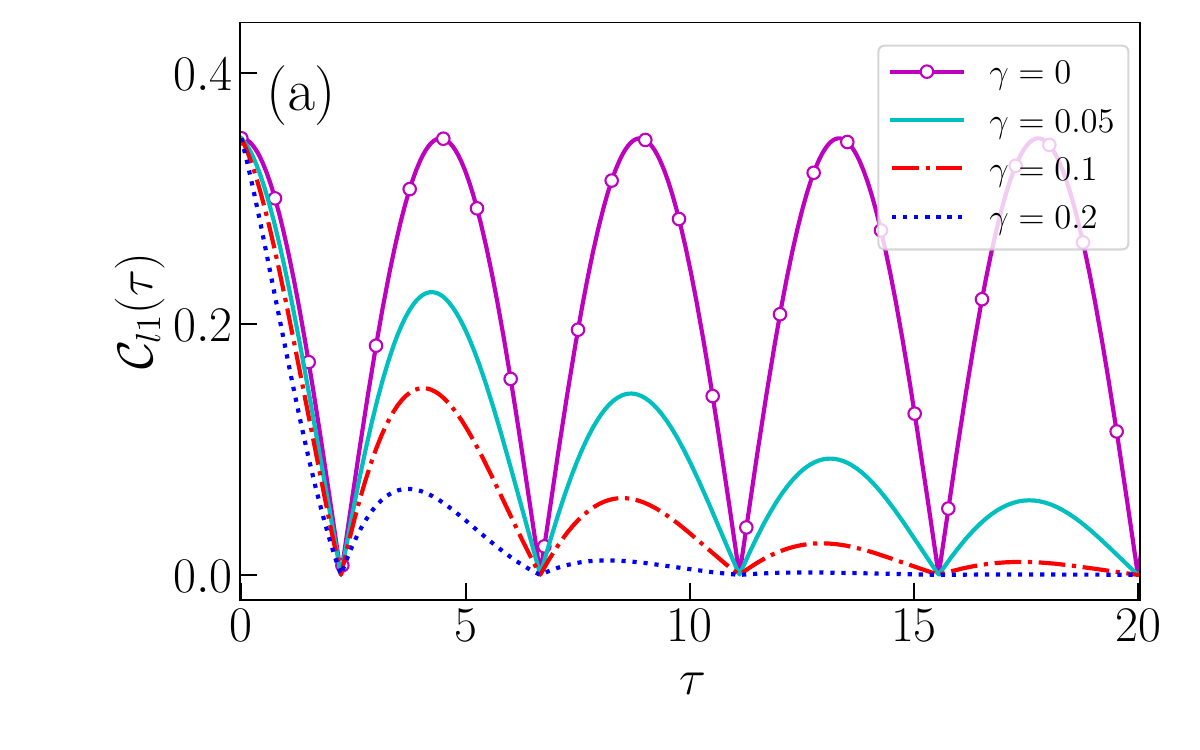}
    \includegraphics[width=0.40\textwidth,height=125px]{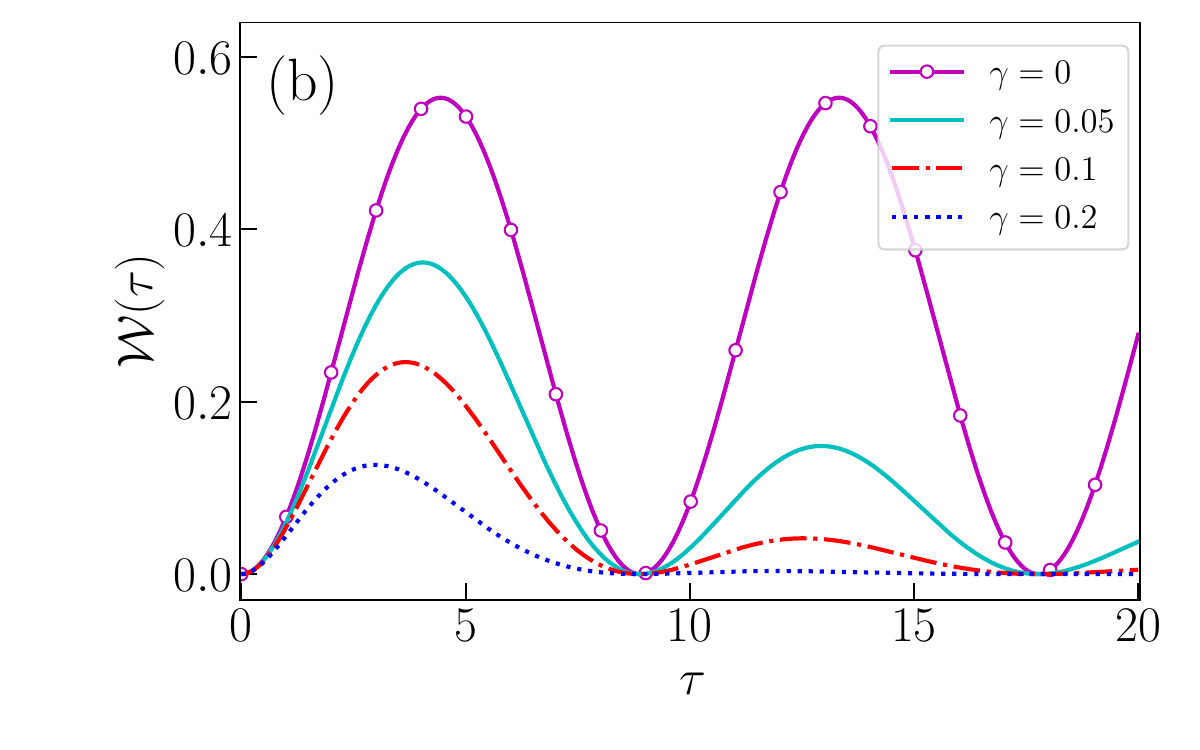}
    \caption{Temporal evolution of (a) $l_1$- norm of coherence and (b) Ergotropy of QB under dephasing environment for different dissipation rates with parameters being set $\theta=\pi/4$,    $g1=g2=2$ and $B=\tilde{D}=\tilde{J}=\tilde{\Omega}=\tilde{T}=1$. }
    \label{fig12}
\end{figure}

Figure (\ref{fig12}) illustrates the temporal evolution of the $l_1-$norm of coherence and ergotropy of the hybrid quantum battery under a dephasing environment for different dephasing strengths. In the absence of dephasing $\gamma=0$, the coherence exhibits sustained periodic oscillations arising from the coherent unitary charging protocol. As the dephasing rate increases, the oscillation amplitude decreases progressively because environmental interactions suppress the quantum superposition generated during the charging process. For sufficiently strong dephasing, the coherence decays rapidly toward zero, indicating that the battery approaches an incoherent mixed state in which the quantum advantage associated with coherent superposition is effectively lost. The open-system results show that increasing dephasing suppresses quantum coherence and modifies the ergotropy dynamics. In particular, coherence and ergotropy do not necessarily evolve synchronously, with the maximum ergotropy occurring even when the coherence is relatively small. This behavior indicates that the relationship between coherence and energy-storage performance is nontrivial under decoherence and should not be interpreted as evidence of a direct causal dependence. The results therefore characterize the sensitivity of both nonclassicality and battery-performance indicators to environmental dephasing.



\end{document}